\documentclass[aps,prb,twocolumn,superscriptaddress,showpacs]{revtex4-2}
\usepackage{graphicx,bm,color,float,textcomp,mathtools,times,gensymb}
\newcommand\esi{$\rm Er_2Si_2O_7$}
\newcommand\ysi{$\rm Yb_2Si_2O_7$}
\newcommand\afm{antiferromagnetic}

\usepackage{siunitx,multirow,nicefrac}
\DeclareSIUnit{\oersted}{Oe}
\DeclareSIUnit\wn{\cm\tothe{-1}}
\DeclareSIUnit{\gauss}{G}
\DeclareSIUnit\electronvolt{e\kern-.12em V}

\newcommand{\vect}[1]{\ensuremath{\boldsymbol{#1}}}

\begin{document}
\title{Magnetic structure, excitations and field induced transitions in the honeycomb lattice $\mathbf{Er_2Si_2O_7}$}
\author{M.~Islam}			\email{Manisha.Islam@warwick.ac.uk} \affiliation{Department of Physics, University of Warwick, Coventry CV4 7AL, United Kingdom}
\author{N.~d'Ambrumenil}	\affiliation{Department of Physics, University of Warwick, Coventry CV4 7AL, United Kingdom}
\author{D.~D.~Khalyavin} 
\affiliation{ISIS Facility, STFC Rutherford Appleton Laboratory, Chilton, Didcot, OX11 0QX, United Kingdom}
\author{P.~Manuel}
\affiliation{ISIS Facility, STFC Rutherford Appleton Laboratory, Chilton, Didcot, OX11 0QX, United Kingdom}
\author{F.~Orlandi}
\affiliation{ISIS Facility, STFC Rutherford Appleton Laboratory, Chilton, Didcot, OX11 0QX, United Kingdom}
\author{J.~Ollivier}			\affiliation{Institut Laue-Langevin, 71 Avenue des Martyrs, CS 20156, 38042 Grenoble Cedex 9, France}
\author{M.~Ciomaga Hatnean}
\altaffiliation[Current affiliations: ]{Materials Discovery Laboratory, Department of Materials, ETH Zurich, 8093 Zurich  and Laboratory for Multiscale materials eXperiments, Paul Scherrer Institute (PSI), 5232 Villigen, Switzerland}
\affiliation{Department of Physics, University of Warwick, Coventry CV4 7AL, United Kingdom}
\author{G.~Balakrishnan}		\affiliation{Department of Physics, University of Warwick, Coventry CV4 7AL, United Kingdom}
\author{O.~A.~Petrenko}		\email{O.Petrenko@warwick.ac.uk} \affiliation{Department of Physics, University of Warwick, Coventry CV4 7AL, United Kingdom}

\date{\today}
\begin{abstract}
We investigate the magnetic properties of the monoclinic D-type \esi\ with a distorted honeycomb lattice using powder and single crystal neutron scattering techniques, as well as single crystal magnetisation measurements.
The powder neutron diffraction shows that below the ordering temperature, $T_{\rm N}=1.85$~K, the compound forms a ${\bf q}=0$ \afm\ structure with four sublattices.
For $H \! \parallel \! a$, magnetisation measurements reveal a narrow, but clearly visible plateau at one third of the magnetisation saturation value.
The plateau's stabilisation is accompanied by a significant increase of the magnetic unit cell, as the magnetic peaks with fractional indices are observed in single crystal neutron diffraction experiments.
At low-temperatures, the inelastic neutron scattering measurements reveal the presence of low-energy dispersionless excitations.
Their spectrum is sensitive to the applied field, it significantly softens on the magnetisation plateau, and demonstrates the behaviour expected for a non-collinear Ising antiferromagnet away from the plateau.  
\end{abstract}
\maketitle
\section{Introduction}		\label{sec_Intro}

The family of rare-earth (RE) disilicates $\rm RE_2Si_2O_7$ is renowned for its complex polymorphism with at least seven different crystal structures possible depending on the particular rare-earth ion, temperature range and pressure during synthesis~\cite{Ito_1968,Felsche_1970,Maqsood_1979}.
Not much is known about the magnetic properties of this family of compounds, however, in several of the polymorphic crystal structures the magnetic RE ions are arranged into a (distorted) honeycomb lattice which makes them potentially very interesting from the contemporary magnetism point of view, particularly when searching for the experimental realisations of the Kitaev model~\cite{Kitaev_2006}.
\ysi, for example, is considered to be a strongly spin-orbit coupled quantum dimer magnet~\cite{Hester_2019}.

\begin{figure}[tb] 
\includegraphics[width=0.9\columnwidth]{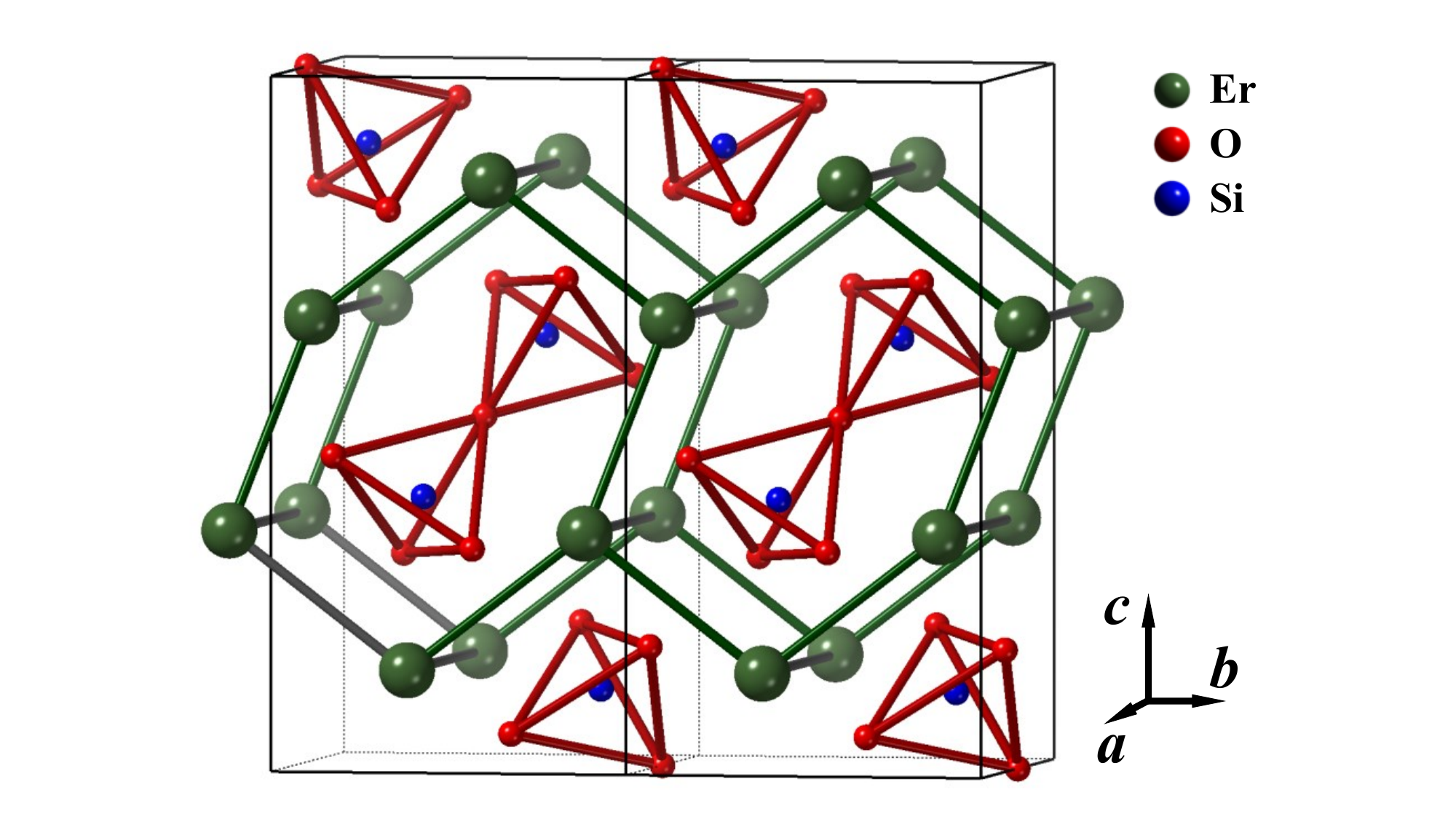}
\caption{Crystal structure of the monoclinic D-type \esi\ (space group $P2_1/b$).
The bonds between the magnetic Er$^{3+}$ ions emphasise the formation of the distorted honeycomb layers stacked along the $a$~axis.
The unit cells are shown in black.}
\label{Fig1_structure}
\end{figure}

Erbium disilicate, \esi, can be formed in three polymorphs, a triclinic $P\Bar{1}$ low temperature phase (labeled as B-type), a monoclinic $C2$/$m$ phase (C-type)~\cite{Hester_2021_C}, and a high temperature monoclinic $P2_1/b$ phase (D-type).
Here we report on the magnetic properties of the monoclinic D-type phase (see Fig.~\ref{Fig1_structure}) and focus on its rather unusual in-field behaviour.
The initial magnetic properties studies reported by Leask {\it et al.}~\cite{Leask_1986} on the D-phase of \esi\ suggested a four-sublattice \afm\ ground state below the ordering temperature of 1.9~K.
This hypothesis has recently received a direct experimental confirmation from powder neutron diffraction measurements~\cite{Hester_2021_D}. 
A pronounced Ising-like character of \esi\ is evident from the highly anisotropic magnetisation curves which were (at least partially) understood within a fairly simple model backed up by Monte Carlo simulations~\cite{Leask_1986}.
However, one experimental observation, a stabilisation of the $\frac{1}{3}$ magnetisation plateau for a field applied along the $a$~axis, was at odds with the theoretical predictions for a model with four sublattices in the magnetic unit cell.
A possibility of the enlarged magnetic unit cell was briefly mentioned, but dismissed as unlikely~\cite{Leask_1986}.
We find that in the narrow region of applied magnetic fields corresponding to the $\frac{1}{3}$ magnetisation plateau, the magnetic unit cell is indeed increased severalfold, as witnessed by a spectacular transformation of the single crystal neutron diffraction patterns.

The appearance of the $\frac{1}{3}$ magnetisation plateaux in triangular lattice antiferromagnets, quantum and classical, is a common feature, and is often associated with the stabilisation of the {\it up-up-down} (uud) structures in Heisenberg, XY and Ising systems by various mechanisms.
In case of only nearest neighbour (NN) interactions, the honeycomb structure is bipartite and not frustrated, however, the next NN interactions could drive it towards a frustrated regime with an extensive ground state degeneracy.
The resulting behaviour is then typical for highly frustrated magnets with complex and fragile ground states sensitive to minute perturbations.
In many cases, the symmetry breaking by an external magnetic field is not trivial and the fractional magnetisation plateaux are expected.  
Recent reports suggest, for example, stabilisation of the $\frac{1}{3}$ and $\frac{2}{3}$ magnetisation plateaux in a quantum antiferromagnet $\rm Cu_2(pymca)_3(ClO_4)$ on a distorted honeycomb lattice~\cite{Okutani_2019,Adhikary_2021}.
For another two-dimensional honeycomb system, $\rm DyNi_3Ga_9$, the plateaux are found for every $\frac{1}{6}$ of the saturation value~\cite{Ninomiya_2017}.
A theoretical study~\cite{GomezAlbarracin_r2021} suggests the stabilisation of the $\frac{1}{3}$ and $\frac{2}{3}$ plateaux in the XY model with third NN interactions on the honeycomb lattice through the order-by-disorder mechanism.
In this paper, we explore the complex magnetisation process in the distorted honeycomb lattice antiferromagnet \esi\ with four non-collinear Ising-like classical spins in the unit cell. 
\section{Experimental procedures} \label{sec_methods}
An \esi\ crystal boule was prepared by the floating zone method using a two-mirror halogen lamp optical image furnace (NEC SC1MDH-11020, Canon Machinery Incorporated).
The growth was performed in air, at ambient pressure and at growth speed of 12~mm/h.
The crystal growth of \esi\ is described elsewhere in a more detailed paper~\cite{Ciomaga_2020}.
Phase purity analysis was carried out on a ground crystal piece of \esi, and the powder X-ray diffraction confirms that the crystallographic structure of the crystal boule is D-type.
As previously reported~\cite{Nair_2019,Ciomaga_2020}, the conventional synthesis by the solid-state method of polycrystalline samples of D-type \esi\ results in the presence of a small amount of an $\rm Er_2SiO_5$ impurity.
To ensure the purity of the \esi\ powder specimen used in our experiments, a polycrystalline sample ($\approx 2$~g) was prepared by crushing a fragment of the crystal boule.

We used a Quantum Design MPMS SQuID magnetometer for the magnetisation measurements in applied magnetic fields of up to 70 kOe.
An iQuantum $^3$He insert~\cite{Shirakawa_2004} allowed the temperature range explored to be extended down to 0.48~K.
In order to estimate the demagnetisation factors, we used a rectangular prism approximation~\cite{Aharoni_1998}.  

The WISH time-of-flight diffractometer~\cite{WISH} at the ISIS facility at the Rutherford Appleton Laboratory (STFC, UK) was used for both the powder (PND) and single crystal neutron diffraction experiments~\cite{2DOIWISH}.
A $^3$He absorption refrigerator provided a base temperature of 0.24~K for neutron diffraction measurements.
For PND, the data obtained at $T=10$~K were used as a background to isolate the magnetic signal for each temperature. Rietveld refinements were performed using the FullProf software suite~\cite{Rod1993}. For the single crystal diffraction measurements, the sample was aligned with its $a$~axis vertical (parallel to the applied field) for access to the reflections within $\pm~7.5$~degrees of the horizontal $b^\star$-$c^\star$ scattering plane.

The inelastic neutron scattering (INS) measurements~\cite{DOIin5} were performed on the time-of-flight IN5 spectrometer~\cite{ollivier2011} at the Institut Laue-Langevin (ILL), Grenoble, France. An incident wavelength of $\lambda = 4.8$~{\AA} ($E_i=3.55$~meV) with an elastic resolution of $\delta E \simeq 86$~$\mu$eV full width at half maximum has been chosen throughout the measurements. For each field and temperature condition, scans between 4 and 10 minutes have been recorded every 1$^{\circ}$ crystal rotation. Data have been reduced under Mantid~\cite{Arnold2014} and analysed with the Horace~\cite{Ewings2016} library under Matlab. 

\section{Results and discussion}	\label{sec_results}
\subsection{Powder neutron diffraction}	\label{sec_PND}
\begin{figure}[tb] 
\includegraphics[width=0.85\columnwidth]{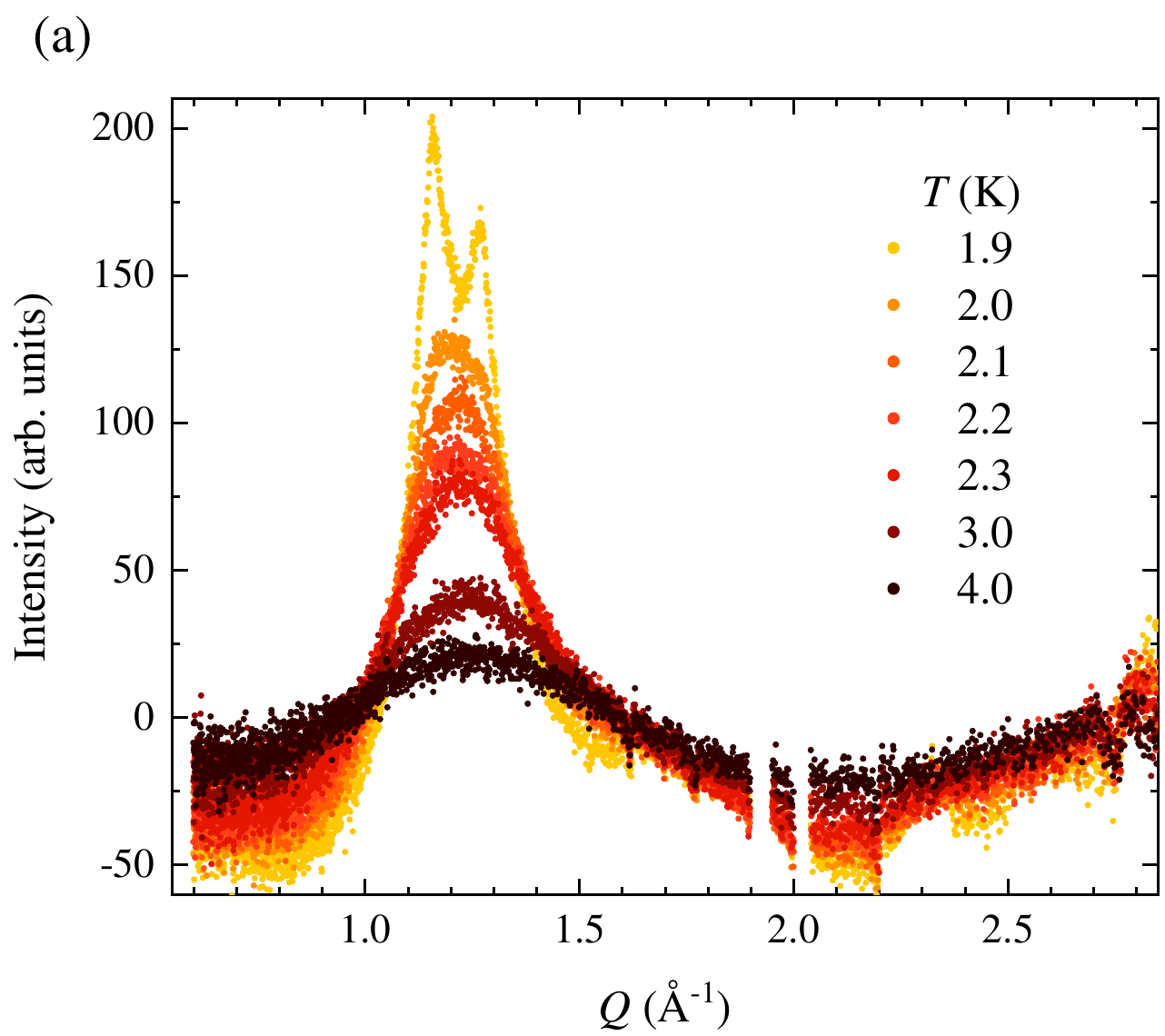}
\includegraphics[width=0.9\columnwidth]{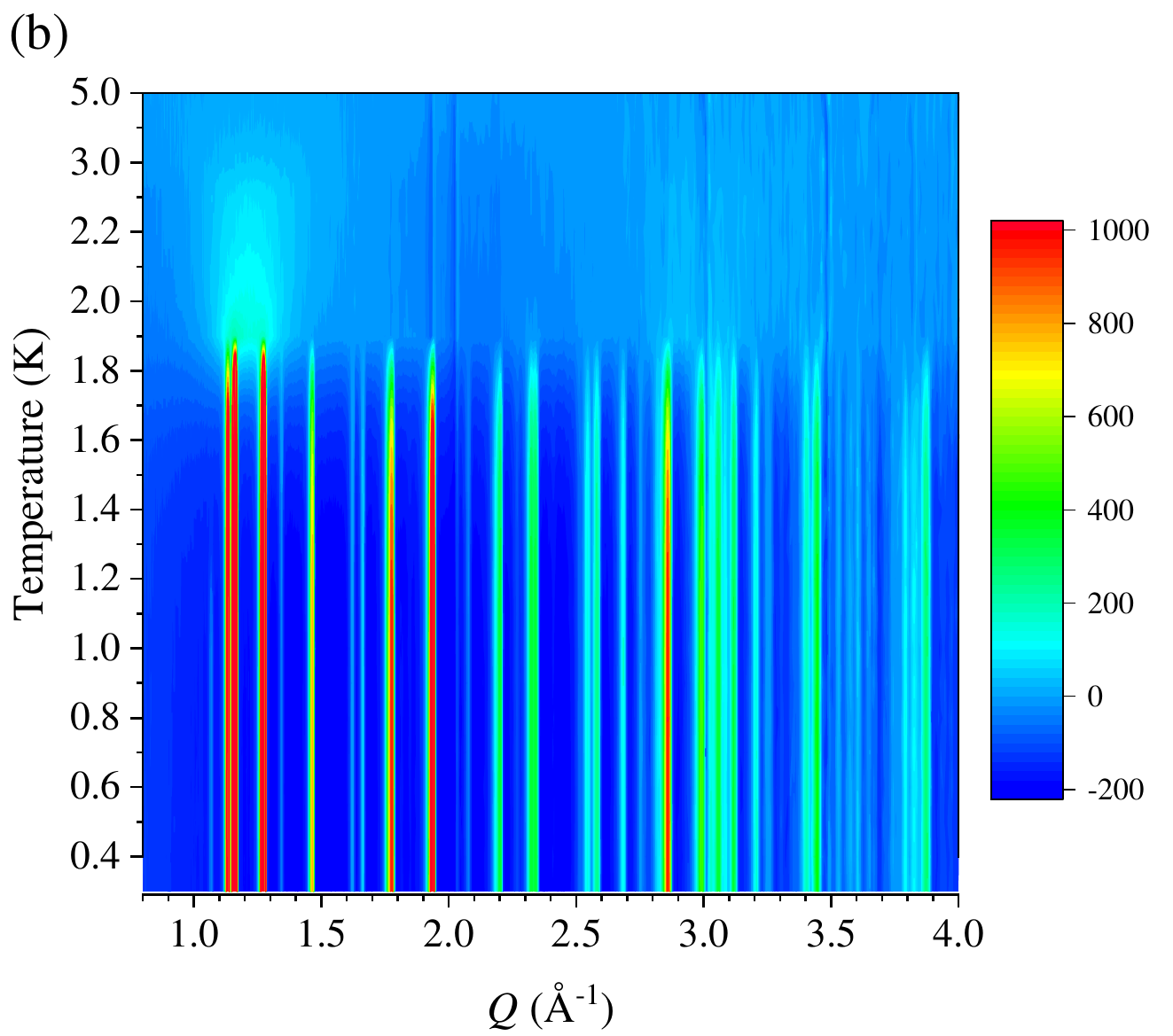}
\caption{Temperature evolution of the magnetic neutron diffraction intensity measured on WISH diffractometer~\cite{WISH} using a powder sample of D-type \esi. Panel (a) shows the development of diffuse signal above ordering temperature. Panel (b) combines all the diffraction data between 0.24 and 5.0~K into an intensity colour map. For both panels, the magnetic contribution is obtained by subtracting a 10~K background.
The omitted regions on panel (a) correspond to the very strong nuclear Bragg peaks for which the subtraction is imperfect.}
\label{Fig2_WISH_powder}
\end{figure}

Fig.~\ref{Fig2_WISH_powder}(a) shows the development of the magnetic correlations in \esi\ upon cooling from 4.0~K to just above the magnetic ordering temperature.
At higher temperatures, the scattering signal is broad, but on approaching the $T_{\rm N}=1.85$~K, it becomes more structured.
The results are consistent with the data presented in~\cite{Hester_2021_D}.
At 1.9~K, two intensity maxima are clearly visible at 1.16 and 1.27~\AA$^{-1}$, the positions corresponding to the strongest magnetic peaks appearing below $T_{\rm N}$. 
Fig.~\ref{Fig2_WISH_powder}(b) emphasises the stabilisation of the long-range magnetic order below $T_{\rm N}=1.85$~K and the absence of further transitions down to at least 0.24~K.

\begin{figure}[tb] 
\includegraphics[width=0.9\columnwidth]{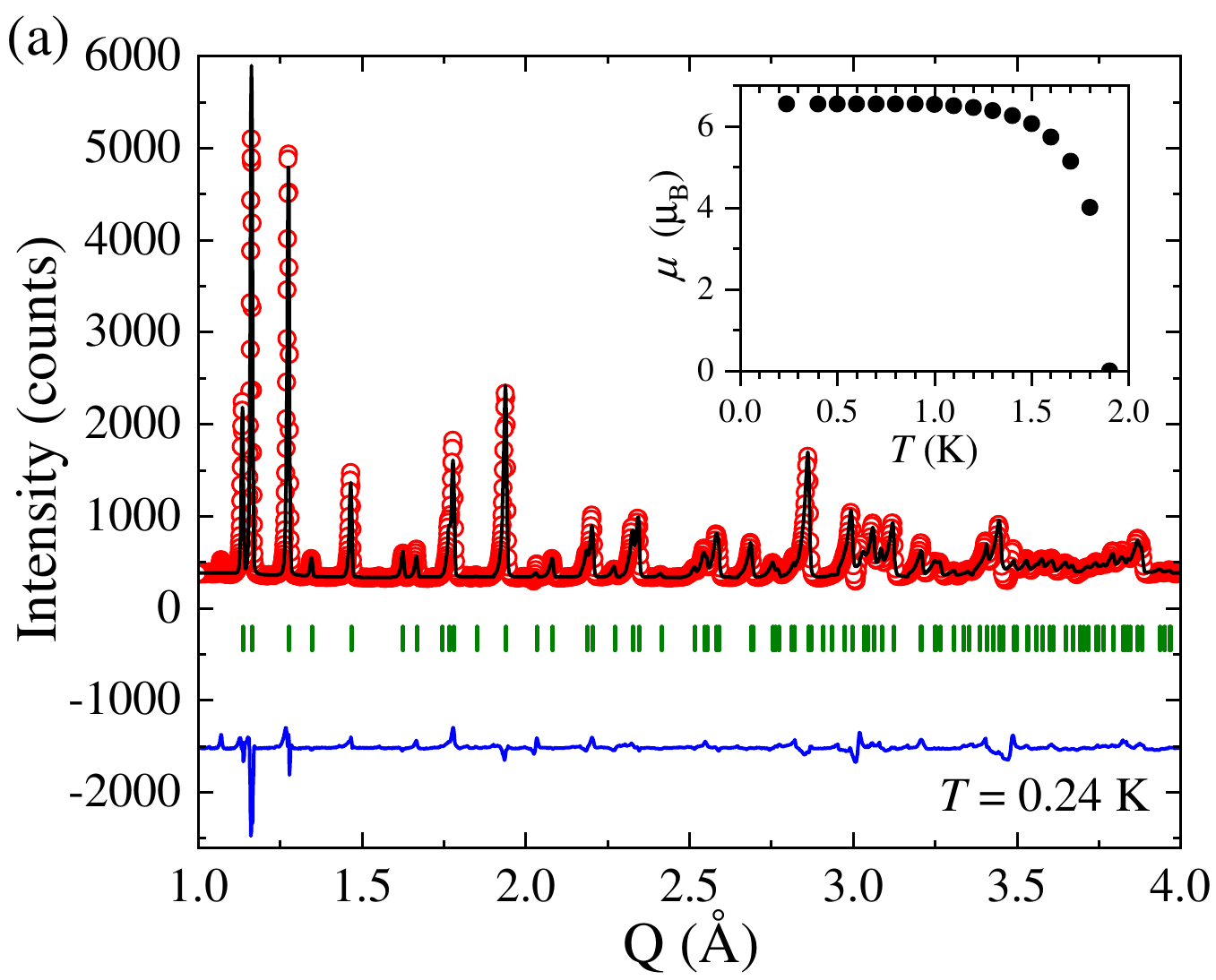}
\includegraphics[width=0.9\columnwidth]{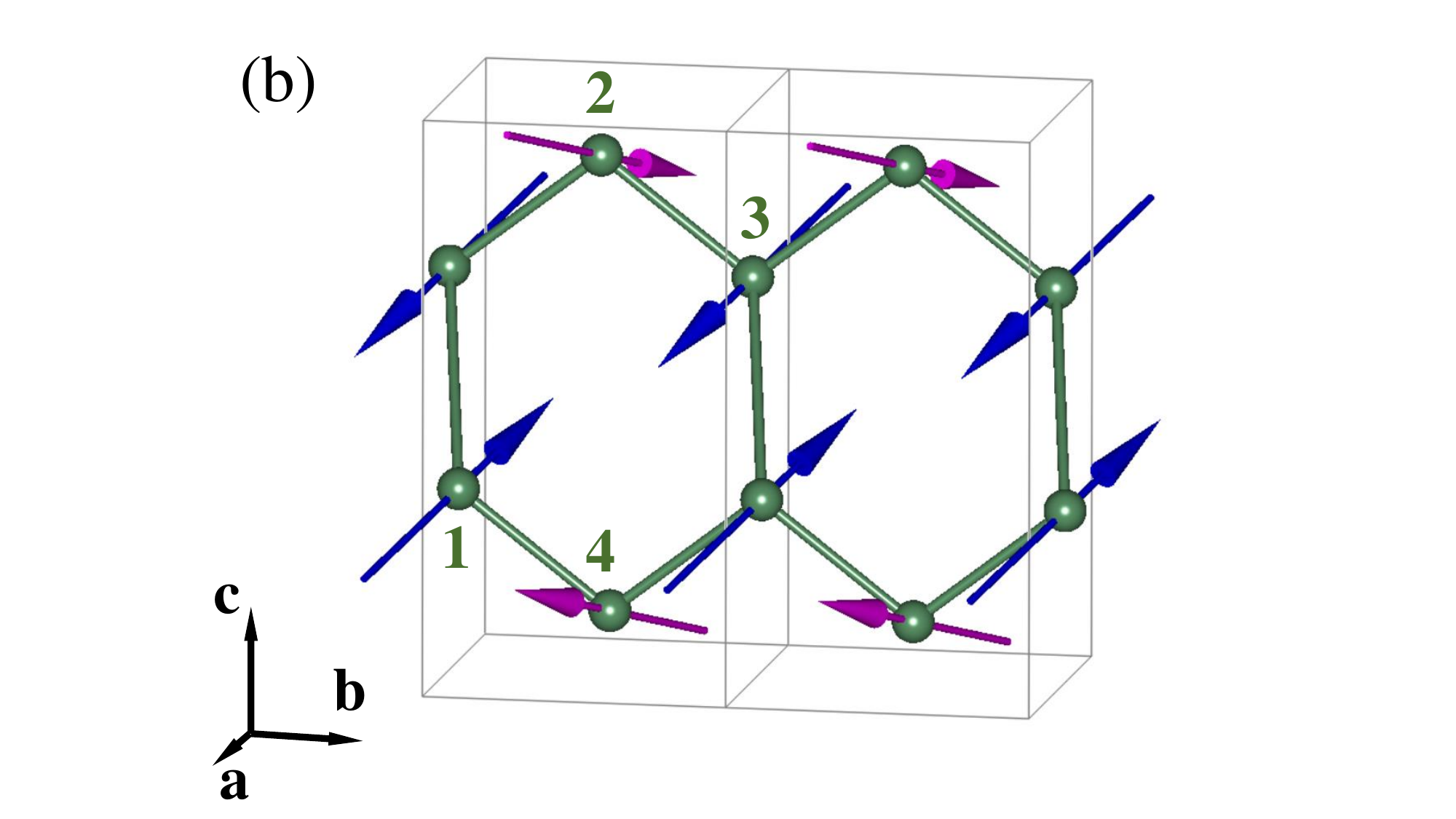}
\caption{(a) Rietveld refinement of the PND data in zero field at 0.24~K.
The background subtracted data (red circles) from banks 2 and 9 of the WISH diffractometer~\cite{WISH} are shown together with calculated pattern (black line) and the difference curve (blue line). Green ticks represent the positions of the magnetic Bragg peaks. The inset follows the temperature dependence of the magnitude of the magnetic moment, $\mu$, of the Er$^{3+}$ ions determined by the refinement.
(b) Magnetic structure of D-type \esi\ obtained from the PND refinement as viewed along the $a$~axis. Unit cell boundaries are shown in black, with the four constituent magnetic ions labelled from 1 to 4 in green.
The positions of the Er$^{3+}$ ions and their moments are listed in Table~\ref{tab:Positions+Moments}.}
\label{Fig3_Mag_structure}
\end{figure}

Refinement of the powder neutron diffraction data taken on WISH at 10~K confirm the monoclinic $P2_{1}/b$ crystallographic space group with lattice parameters of  $a$~=~4.6932~\AA, $b$~=~5.5586~\AA, $c$~=~10.7942~\AA, $\alpha$~=~90~$^{\circ}$, $\beta$~=~90~$^{\circ}$, $\gamma$~=~96~$^{\circ}$. The magnetic contribution of the data at 0.24~K are shown in Fig.~\ref{Fig3_Mag_structure}(a), after the subtraction of the 10~K measurement in the paramagnetic phase.
The presence of magnetic intensities on top of the nuclear Bragg peaks indicates $\textbf{q}=0$ propagation vector. The magnetic structure solution was approached based on the space group representation (irrep) theory~\cite{Izyumov1991}, assuming an irreducible nature of the magnetic order parameter. We focus on the data collected in the detector banks 2 and 9. Systematic testing of the relevant irreps in the refinement of the diffraction pattern revealed that only mGM2-~\cite{Stokes2022, Campbell2006} provided good quality fitting ($R_{\rm magnetic}=10.9 \%$). mGM2- is a one-dimensional irrep entering three times into decomposition of the reducible representation associated with the Er Wyckoff position $4e$.
The corresponding three basis vectors transform the $\mu_a$, $\mu_b$ and $\mu_c$ components of the ordered moment specified in Table I. All three components are non-zero, yielding a non-collinear magnetic structure shown in Fig.~\ref{Fig3_Mag_structure}b with a total moment of 6.55~$\mu_{\rm B}$ at $T=0.24$~K.
A magnetic structure with the unit cell containing four magnetic ions at unequivalent positions was refined, with all magnetic contributions found only at the nuclear peak positions and corresponding to the propagation vector $\boldsymbol{q} = 0$. A fit with $R~=~22.6~\%$, $wR~=~21.7~\%$, and $R_{Bragg}~=~0.9~\%$ was achieved for $T=0.24$~K. The temperature dependence of the magnitude of the moment of the Er$^{3+}$ ions is shown in the inset. The moment remains practically independent of temperature at 6.55~$\mu_{\rm B}$ until 1~K, above which it decreases with temperature and drops to zero at 1.9~K.

Fig.~\ref{Fig3_Mag_structure}(b) illustrates the resulting magnetic structure of \esi\ after Rietveld refinement.
The magnetic ions, shown in green, present a distorted honeycomb arrangement.
The unit cell contains four magnetic Er$^{3+}$ ions, their positions are labelled in  Fig.~\ref{Fig3_Mag_structure}(b). The Er$^{3+}$ moments at positions 1 and 3 (blue arrows), and 2 and 4 (pink arrows) are \afm ally aligned to each other and share an easy magnetisation axis.
Table~\ref{tab:Positions+Moments} details the positions of the ions and components of the magnetic moments as well as the average moment per ion when the spins are flipped by the applied field. The magnetic structure implies $P2_{1}'/c$ symmetry with the unit cell related to the paramagnetic structure as (-1,0,0), (0,0,-1) and (0,-1,0) (see the supplementary mcif file~\cite{supplementary} for details).

\begin{table}[tb]
\centering
 \begin{tabular}{c c c c c r r r c c} 
 \hline \hline
  Ion & \multicolumn{3}{c}{Positions} & & \multicolumn{1}{c}{$\mu_a$} &  \multicolumn{1}{c}{$\mu_b$} &  \multicolumn{1}{c}{$\mu_c$} & & Moment \\
  & \multicolumn{3}{c}{$(\vect{a}, \vect{b}, \vect{c})$ basis} & & \multicolumn{3}{c}{$[\mu_{\rm B}]$} & & $[\mu_{\rm B}]$ \\ \hline
 1 & 0.89 & 0.10 &  0.35 &  \;\; & -5.62 &  2.11 &  2.08 & \; \; & 6.55 \\ 
 2 & 0.11 & 0.40 &  0.85 & & -5.62 &  2.11 & -2.08 & & 6.55 \\
 3 & 0.11 & 0.90 &  0.65 & &  5.62 & -2.11 & -2.08 & & 6.55 \\
 4 & 0.89 & 0.60 &  0.15 & &  5.62 & -2.11 &  2.08 & & 6.55 \\
 \hline  \hline
 $\langle\vect{m}_a\rangle$ & & & & &  5.62 & -2.11 & 0.00  & & 6.20 \\ 
 $\langle\vect{m}_c\rangle$ & & & & &  0.00 &  0.00 & 2.08  & & 2.08\\ \hline\hline
 \end{tabular}
 \caption{Ionic positions and magnetic moments of the four Er$^{3+}$ ions, obtained from the refinement of zero-field PND data taken at 0.24 K.  The labels $\mu_{a,b,c}$ denote the components of the magnetic moments along unit vectors in the $\vect{a}$, $\vect{b}$ and $\vect{c}$ directions and have an error of~$<1~\%$.
 The moments $\langle \vect{m}_a\rangle$ and $\langle \vect{m}_c\rangle$ are the theoretical average moments per ion expected if ions flip to align as far as possible with an applied field in the $a$ and $c$~directions. }
 \label{tab:Positions+Moments}
\end{table}
\subsection{Magnetisation measurements}	\label{sec_mag}
\begin{figure}[tb] 
\includegraphics[width=0.9\columnwidth]{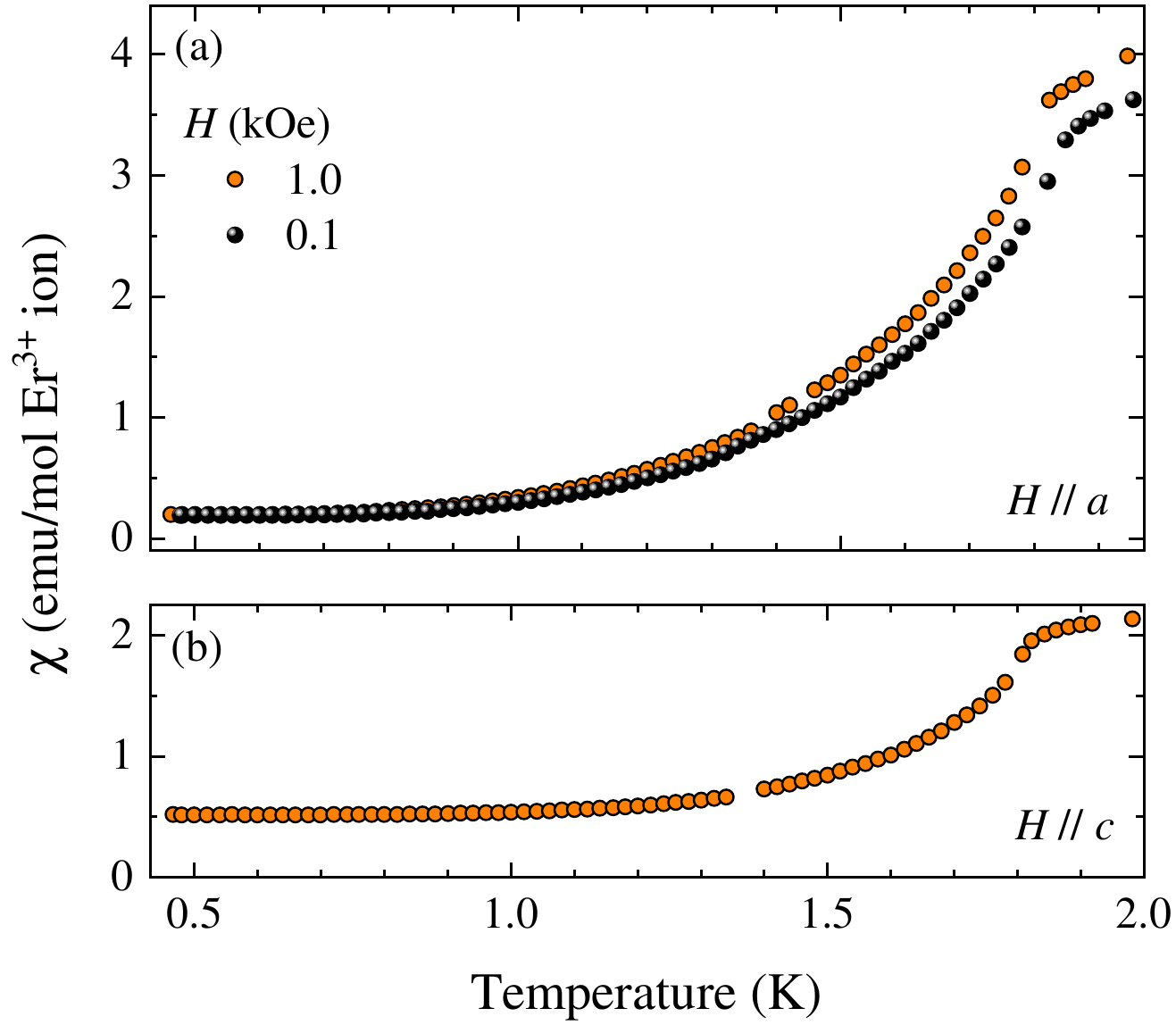}
\caption{(a) Temperature dependence of the susceptibility of D-type \esi\ measured in 0.1 and 1.0~kOe for $H \! \parallel \! a$.
(b) Temperature dependence of the susceptibility measured in 1~kOe for $H \! \parallel \! c$.
For both field directions, the applied field is quoted without taking into account demagnetisation corrections.}
\label{Fig4_chi}
\end{figure}

Fig.~\ref{Fig4_chi}(a) summarises the low-temperature susceptibility measurements in 0.1 and 1.0~kOe along the $a$~axis.
The magnetic ordering temperatures in low fields is seen as a sharp drop in the susceptibility below $T_{\rm N}=1.85$~K.
No appreciable difference has been noticed in the susceptibility measured in field-cooled and in zero-field-cooled conditions for the applied fields of 0.1 and 1.0~kOe. 

\begin{figure}[tb] 
\includegraphics[width=0.9\columnwidth]{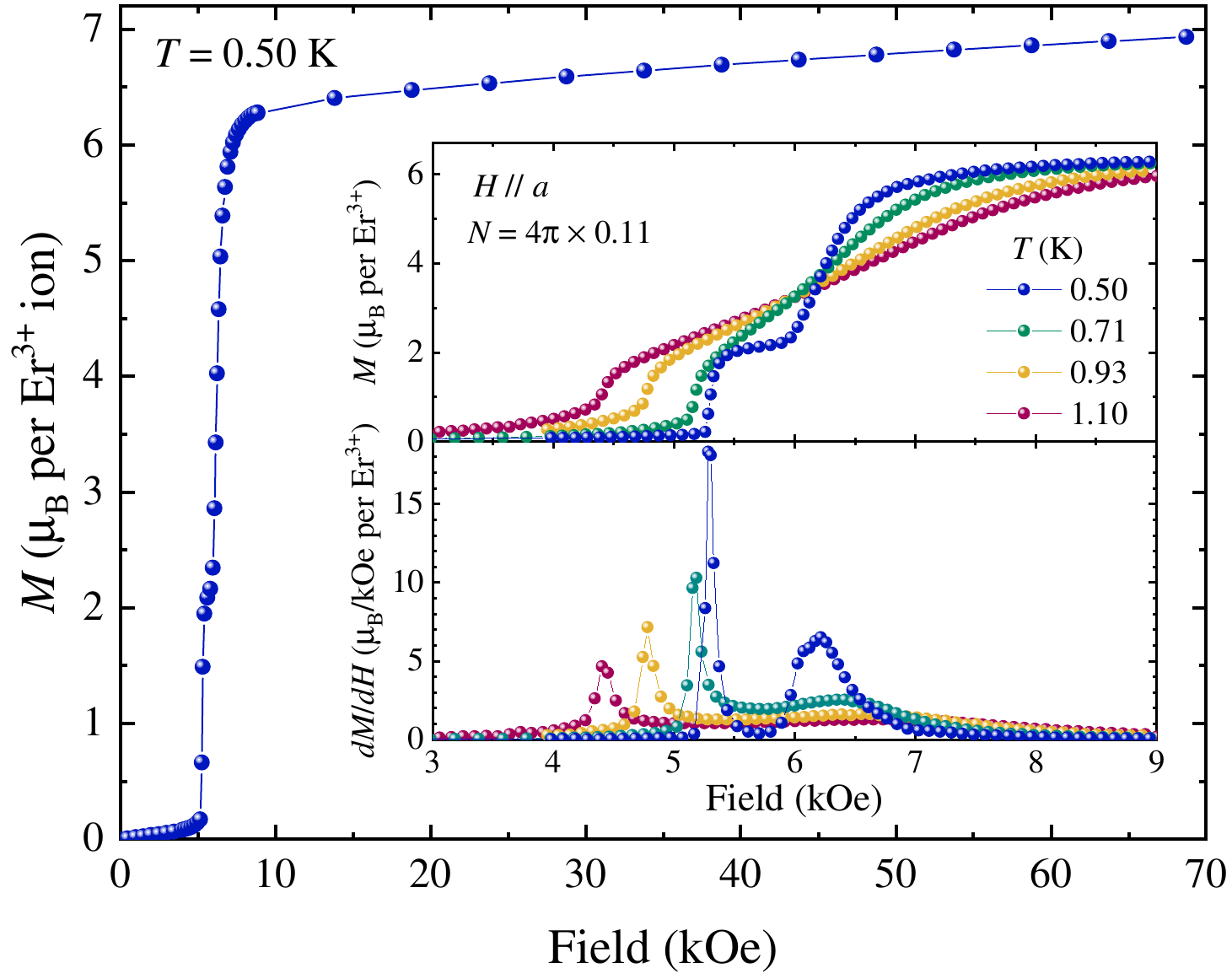}
\caption{Field dependence of the magnetisation of D-type \esi\ measured at 0.5~K for $H \! \parallel \! a$.
The inset focuses on the transition found at 5.3~kOe and follows the temperature evolution of (top panel) the magnetisation, $M$, and (bottom panel) its field derivative $dM/dH$.
The applied field is corrected using the demagnetisation factor $N_a=0.11 \times 4\pi$.}
\label{Fig5_Mag_a}
\end{figure}

\begin{figure}[t] 
\includegraphics[width=0.9\columnwidth]{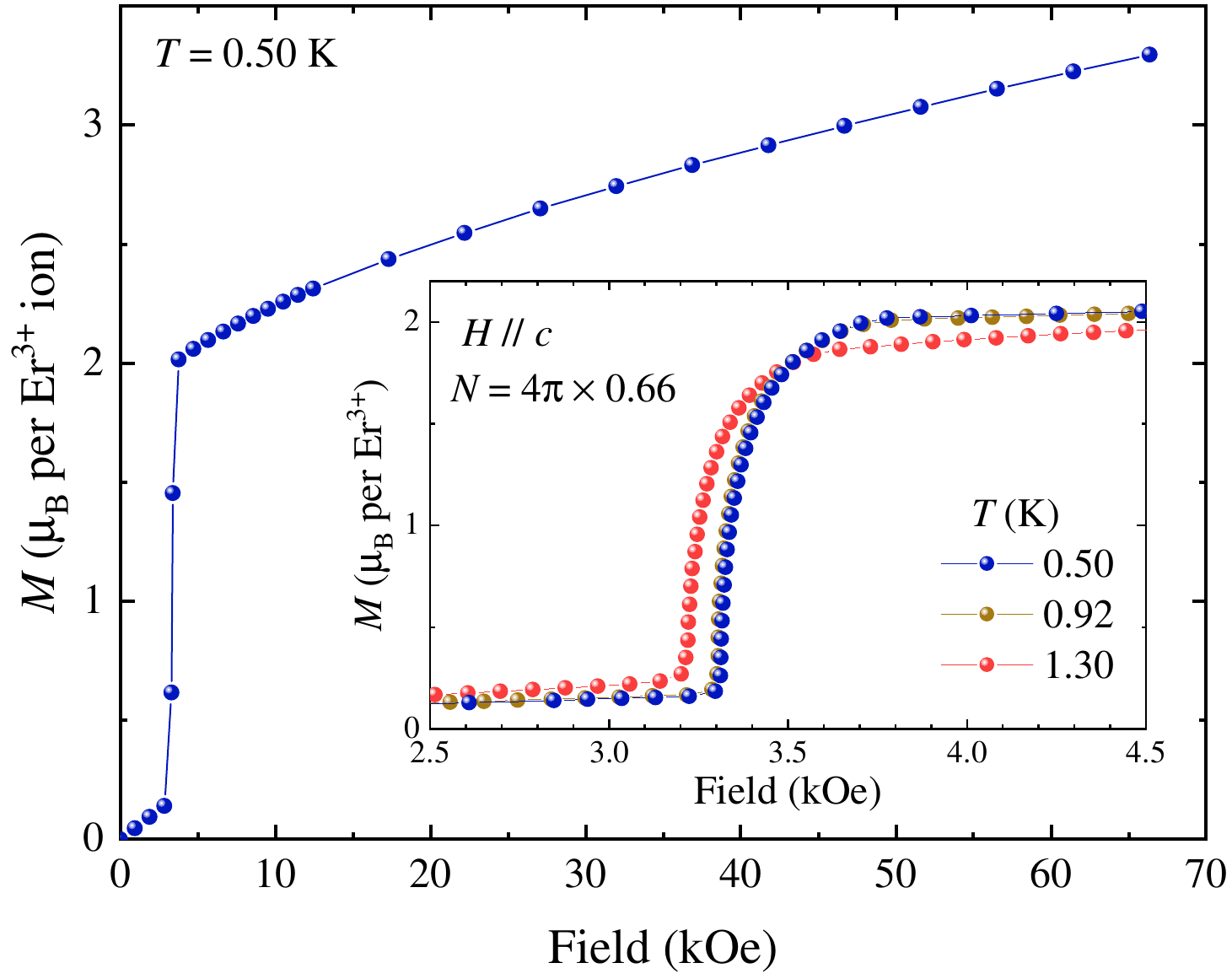}
\caption{Field dependence of the magnetisation of D-type \esi\ measured at 0.5~K for $H \! \parallel \! c$.
The inset shows the transition observed at 3.3~kOe on an expanded scale and shows the field dependence of the magnetisation around the transition at different temperatures.
The applied field is corrected using the demagnetisation factor $N_c=0.66 \times 4\pi$.}
\label{Fig6_Mag_c}
\end{figure}

Figs.~\ref{Fig5_Mag_a} and \ref{Fig6_Mag_c} illustrate the magnetisation process in \esi\ for two directions of an applied magnetic field, $H \! \parallel \! a$ and $H \! \parallel \! c$.
For both field directions, a rather sharp increase in magnetisation $M(H)$ is observed in relatively weak fields, 5.3~kOe for the field applied along the $a$~axis and 3.3~kOe for the field along the $c$~axis.

For $H \! \parallel \! a$, at the lowest temperature, a clear plateau in magnetisation is visible, extending from 5.3 to 6.0~kOe.
Above the plateau, the $M_{H  \parallel  a}$ increases again rather sharply reaching 5.9~$\mu_{\rm B}$ per Er$^{3+}$ ion already in a field of 7~kOe and 6.5~$\mu_{\rm B}$ in a field of 20~kOe.
This value is in agreement with the total moment of 6.55~${\mu}_{\rm B}$ obtained from the PND refinements (see Fig.~\ref{Fig3_Mag_structure}).
Above 20~kOe, the magnetisation tends to saturate with further field increase up to 70~kOe, resulting in a small increase of $M_{H \parallel a}$ to 6.9~$\mu_{\rm B}$ per Er$^{3+}$ ion.

The transitions to and from the magnetisation plateau are temperature sensitive (see inset in Fig.~\ref{Fig5_Mag_a}) -- a minimum in the field derivative $dM/dH$ becomes barely visible at $T=0.71$~K and disappears completely for $T=0.93$ and 1.1~K despite the temperature remaining well below the $T_{\rm N}$.

For $H \! \parallel \! c$, the transition at 3.3~kOe is accompanied by a sudden jump of magnetisation from 0.1 to 2.1~$\mu_{\rm B}$, no signs of a plateau stabilisation could be detected for this direction of an applied field (see inset in Fig.~\ref{Fig6_Mag_c}).
Above the transition field, the $M_{H  \parallel  c}$ continues to significantly increase reaching 3.3~$\mu_{\rm B}$ per Er$^{3+}$ ion in a maximum field of 70~kOe.
Unlike the case of $H \! \parallel \! a$, the magnetisation process around the critical field for $H \! \parallel \! c$ is not particularly temperature sensitive.

The magnetisation measurements results are in agreement with the previous report~\cite{Leask_1986}.
Small variations in the values of the observed critical fields are most likely caused by the uncertainty in the estimates of the demagnetisation factor, as the sample used was not a perfect rectangular prism.
The estimates of demagnetisation factor of the sample used in magnetisation measurements for $H \! \parallel \! a$ and $H \! \parallel \! c$ are 0.11 and 0.66 (in the units of 4$\pi$) respectively~\cite{Aharoni_1998}.  
Given a relatively high density of \esi\ and large magnetic moments of Er$^{3+}$ ions, the demagnetisation field corrections make an appreciable impact on the values of the transition fields.

\subsection{Single crystal neutron diffraction}	\label{sec_WISH}
\begin{figure*}[tb] 
\includegraphics[width=1.7\columnwidth]{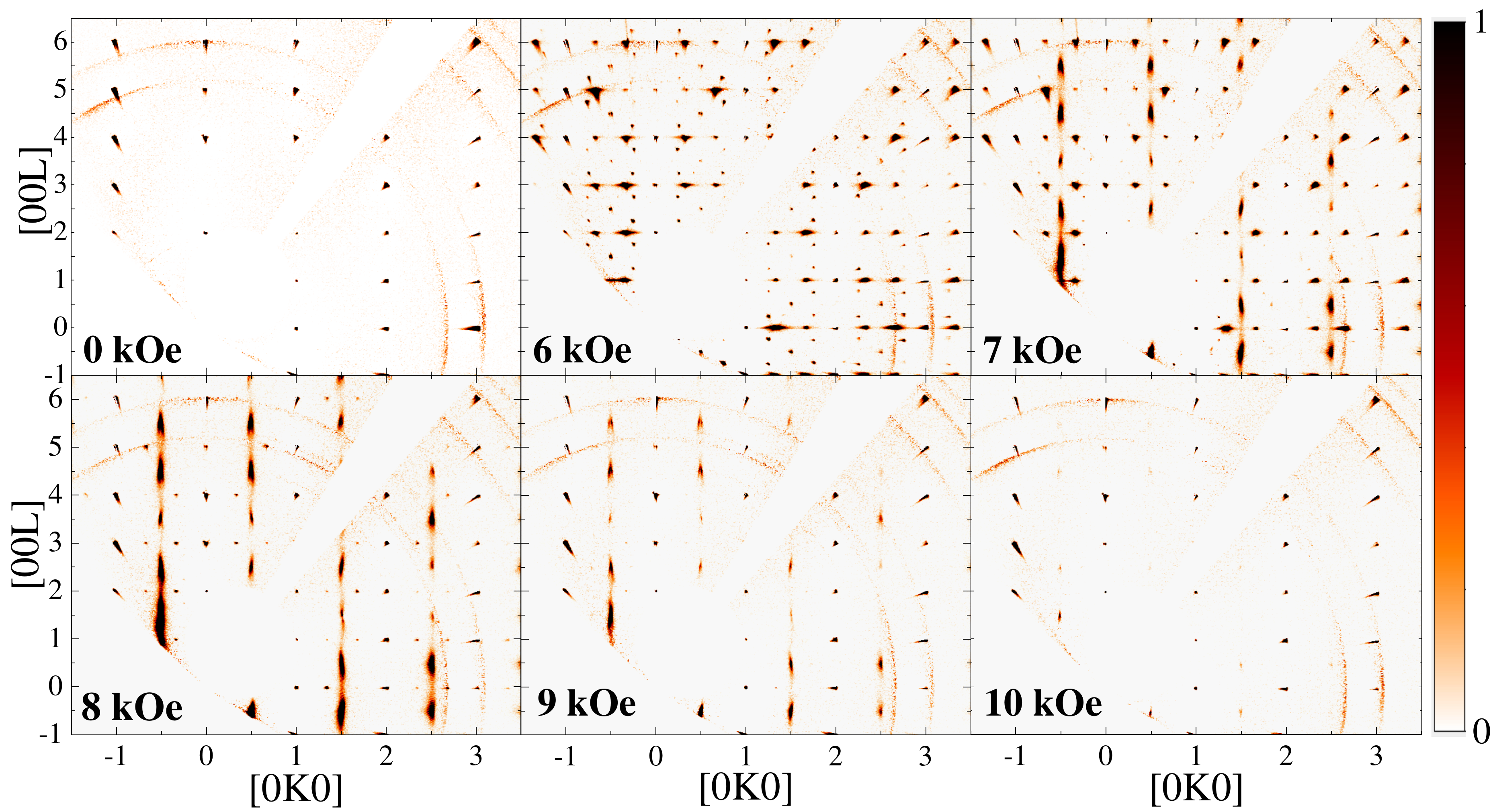}
\caption{Magnetic single crystal neutron diffraction maps of $(0kl)$ plane for D-type \esi\ measured on the WISH diffractometer at $T=0.24$~K in different fields applied along the $a$~axis.
The magnetic signal is obtained by subtracting the $T=2$~K background.
Only sharp Bragg peaks with integer $k$ and $l$ indices are observed in zero field and in applied fields above 10~kOe (see~\cite{supplementary}). The scattering patterns in the fields corresponding to magnetisation plateau are dominated by the appearance of the peaks and diffuse scattering features at fractional positions in reciprocal space.}
\label{Fig7_WISH_Xtal}
\end{figure*}

We followed the magnetisation process in \esi\ for a field applied along the $a$~axis by performing single crystal neutron diffraction measurements on the WISH diffractometer at ISIS~\cite{WISH}.
Above and below the magnetisation plateau region the observed magnetic pattern consists of the ${\bf q}=0$ peaks, while on the plateau, additional signal around the non-integer $(0kl)$ positions becomes clearly visible (see Fig.~\ref{Fig7_WISH_Xtal}). The magnetic signal consists of a mixture of the resolution-limited Bragg peaks and much broader diffuse scattering features. 
There are sharp peaks around the positions described with the propagation vector ${\bf q} = (0 \frac{1}{2} \frac{1}{2})$ at 6~kOe. The intensity profiles of these peaks become broader along the $c^\star$ direction in 7~kOe. In 8~kOe the signal could be described as a collection of diffuse scattering rods with undulating intensity running along the $c^\star$ direction at $\frac{1}{2}$-integer values of $k$.
On further increase of the applied field to 9~kOe the intensities of these rods decrease and they almost disappear at 10~kOe.

The phase at 6~kOe, which we associate with the magnetisation plateau regime,  is also characterised by the appearance of the relatively weak resolution-limited peaks at the ${\bf q} = (0 \frac{1}{4} \frac{1}{4})$ positions. 
These peaks are barely visible at 7~kOe and absent in data taken at the other fields. The diffraction pattern in 6~kOe is also characterised by the presence of broad diffuse features elongated along the $b^\star$ direction at the ${\bf q} = (0 \frac{1}{3} 0)$ positions, which form the undulating intensity rods running along the $b^\star$ axis. These rods are also present in 7~kOe, but in 8~kOe they are transformed into a collection of weak and narrow peaks at  ${\bf q} =(0\frac{1}{3}0)$ positions. 

\begin{figure}[tb] 
\includegraphics[width=0.95\columnwidth]{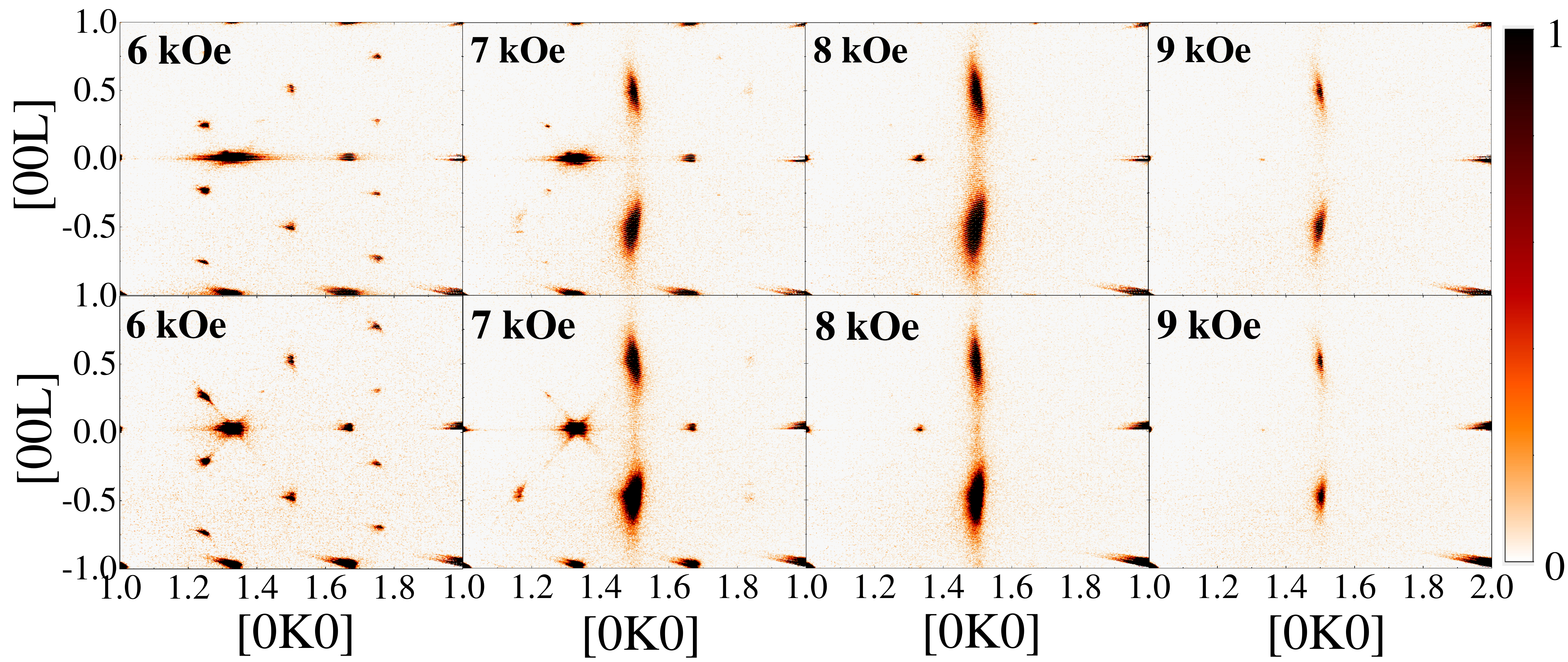}
\caption{Development of the magnetic correlations at 0.24~K in D-type \esi\ in a field from 6 to 9~kOe applied along the $a$~axis before (top row) and after exposure to a higher field of 50~kOe (bottom row).}
\label{Fig8_WISH_Xtal}
\end{figure}

The additional non-integer diffuse features appear as sharp peaks with intensities comparable to the main magnetic reflections at the ${\bf q} = 0$ positions. We have therefore chosen an appropriate scale (2~\% of the intensity of the Bragg reflections at the ${\bf q} = 0$ positions) to best illustrate the weaker diffuse signal within the plateau regime in Figs.~\ref{Fig7_WISH_Xtal} and~\ref{Fig8_WISH_Xtal}. At applied fields of 8 and 9~kOe the undulating diffuse features appear to tilt along $k$ while traversing along $l$. The direction of the tilt alternates as the Brilloiun zone boundaries are crossed with increasing $l$. A similar undulation of diffuse scattering rods has previously been linked to the presence of an incommensurate state in the honeycomb systems SrHo$_{2}$O$_{4}$ and SrEr$_{2}$O$_{4}$~\cite{Young_2013, Hayes_2011}.

The rather irregular shape of the sample, a result of natural cleaving of the as-grown boule, used in the diffraction experiments means that we cannot accurately calculate the demagnetising factor $N$ and a comparison between the neutron diffraction and magnetisation features is not straightforward. However, just for estimation purposes, if we were to assume that $N_a \approx 0.22 \times 4\pi$, then on a plateau, the demagnetising field is around 12\% of the applied field (further discussion is given section~\ref{sec_model}). Therefore for the single crystal diffraction measurements, an applied field of 6~kOe corresponds to the left-hand side of the magnetisation plateau, the field of 7~kOe is very close to its right-hand side, while 8~kOe corresponds to the region of a rapid magnetisation growth above the plateau.
The fields of 9 and 10~kOe would then correspond to a nearly saturated phase.
 

Fig.~\ref{Fig8_WISH_Xtal} compares the magnetic correlations in identical applied fields along the $a$ axis at before (top) and after exposure to a significantly higher field of 50~kOe (bottom). At all fields within the plateau boundaries, there is an apparent asymmetry when looking at the diffuse scattering  patterns in Fig.~\ref{Fig8_WISH_Xtal}. The difference is most pronounced along $b^\star$, when comparing the diffuse features at ${\bf q}=(0\frac{1}{3}0)$ and $(0\frac{2}{3}0)$. The signal at $(0 \frac{2}{3} 0)$ is a distinct, sharp Bragg peak-like feature while at $(0 \frac{1}{3} 0)$, there is a relatively large diffuse scattering feature elongated along $b^\star$. The diffuse scattering signals on the upward and downward field sweep also differ, with well-defined, diagonal lines of intensity traversing through the ${\bf q}= (0 \frac{1}{3} 0)$ and the surrounding  $(0 \frac{1}{4}\frac{1}{4})$ peaks appearing on the downward sweep.
The resulting star-like shapes are much more pronounced in the patterns taken after the sample was subjected to a high applied field, above the saturation field.

A full set of the diffraction maps measured on WISH can be found in the Supplementary Material~\cite{supplementary}.

\subsection{Inelastic neutron scattering}	\label{sec_IN5}
\begin{figure*}[tb] 
\includegraphics[width=1.5\columnwidth]{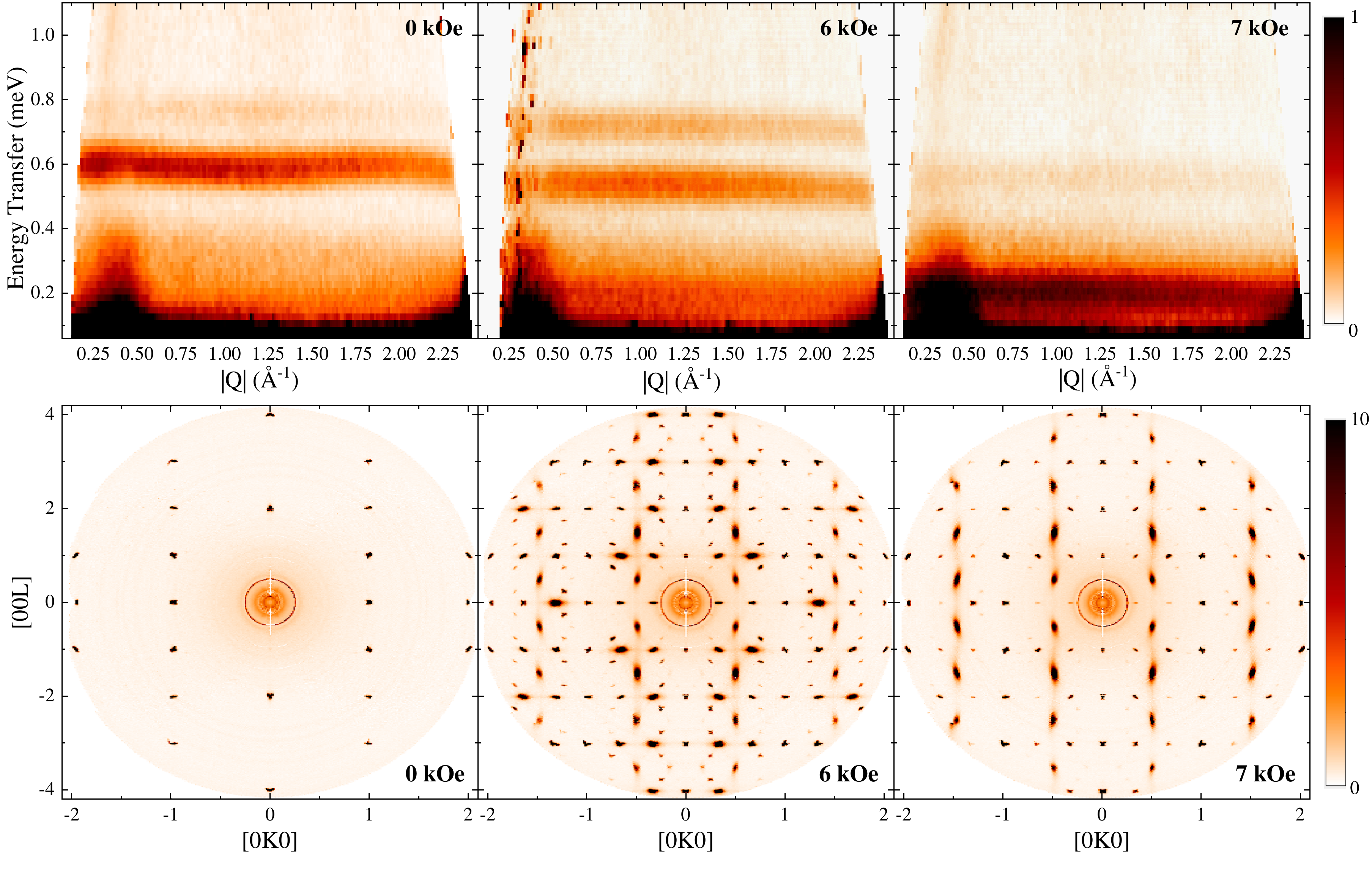}
\caption{Top: Evolution of the dynamical structure
factor $S(Q, E)$ with field taken on the IN5 spectrometer on a single crystal sample of D-type \esi\ at $T=0.045$~K. The resulting energy spectra have been  averaged over the rotation of the crystal due to the presence of dispersionless excitations only.
The sample was rotated around the $a$ axis~$\sim 43^{\circ}$, parallel to an applied field, $H$, while the average is taken over the whole detector range. The high intensity feature at $|Q|=0.35~\rm{\AA^{-1}}$ is an artefact of the cryomagnet and can be ignored.
Bottom: Corresponding symmetrised diffraction maps for the zero-energy transfer in the $(0kl)$ plane.}
\label{Fig9_IN5_maps}
\end{figure*}

The measurements of the excitations performed on the IN5 spectrometer at ILL with a single crystal sample of \esi\ were taken with the vertical magnetic field applied along the $a$~axis.
The measurements of the elastic signal in the $b^{\star}$-$c^{\star}$ scattering plane allowed for an identification of the magnetic state, as a direct comparison with the magnetisation data is again complicated by the irregular sample's shape and the associated uncertainty of the demagnetising factor.
We therefore accompany the INS data with the maps of the elastic intensity observed in the $b^\star$-$c^\star$ plane. Whenever possible, the diffraction intensity maps have been folded according to the crystal symmetries in order to improve the statistics.

For all the measurements made on IN5, only dispersionless excitations were found with no significant dependence of the excitation energy on the scattering vector. We therefore present our measurements as an average over the rotation of the crystal, producing a powder-average spectrum, obtained by grouping the pixellated data of the detector in ($\vert$Q$\vert$, E) bins ie. Debeye-Scherrer cones over the entire detector range.

As the orientation of the crystal only allows for measurement of the $b^\star$-$c^\star$ plane, this method of averaging cannot represent a true powder-average containing the information for all orientations of the crystals with the incident and scattered neutrons spanning all possible crystal symmetry directions. Our case covers the symmetry directions of interest, in which there is no dispersion of the inelastic modes. It should be noted, however, the Q-dependence of the intensity of these excitations cannot be determined.
We therefore present the powder-averaged INS data as colour-coded intensity maps, for different values of the applied field, see Fig.~\ref{Fig9_IN5_maps}. 

In zero field, a high intensity excitation is observed at 0.58~meV, along with a second, lower intensity branch at 0.75~meV, differing from previous powder INS measurements in~\cite{Hester_2021_D}, where two branches were seen at 0.2~meV and 0.6~meV. With increasing field, the higher intensity branch appears to split into two branches at 0.74~meV and 0.43~meV in 2~kOe.
The energies of the two branches increase and decrease linearly as the applied field is increased further. 
By 5~kOe, the energy gap between the branches reaches 0.76~meV.
The sample enters the plateau phase at 5.5~kOe, confirmed by the presence of the non-integer peaks in the elastic reciprocal space maps.
The transition to a plateau regime is accompanied by the appearance of three excitation branches.
A full set of the field dependent excitation spectra can be found in the Supplementary Material~\cite{supplementary}.

\begin{figure}[h!] 
\includegraphics[width=0.79\columnwidth]{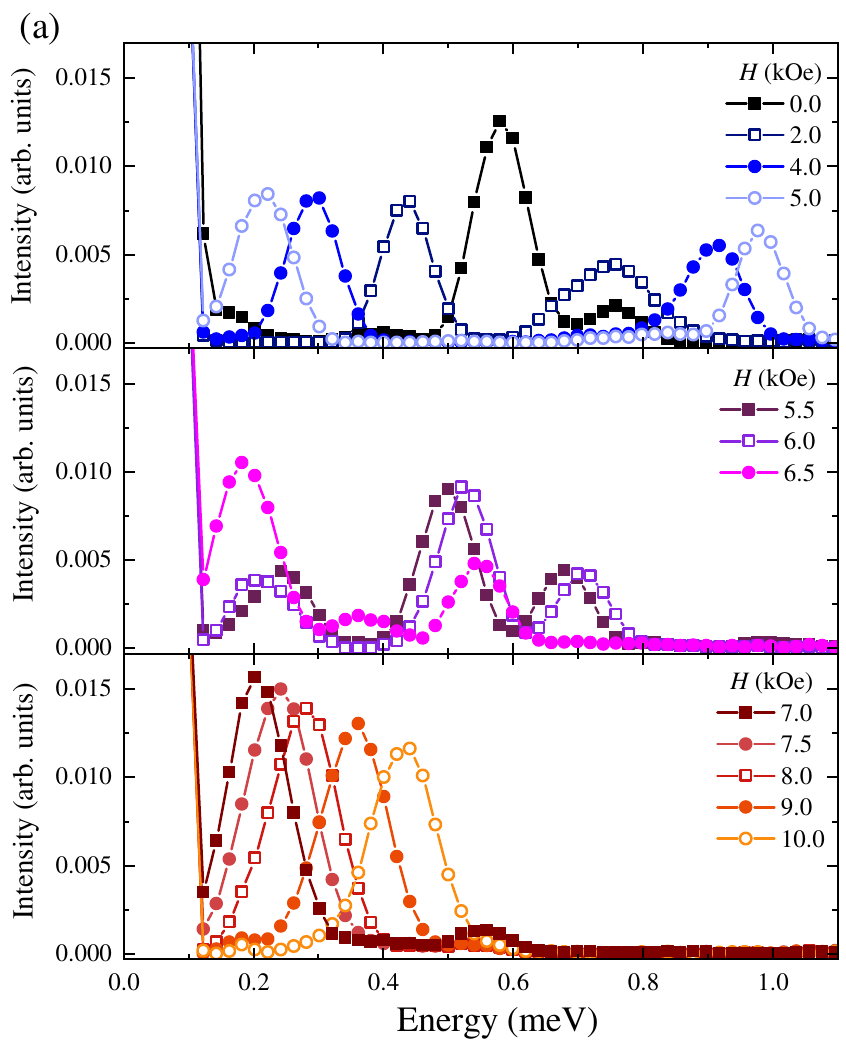}
\includegraphics[width=0.79\columnwidth]{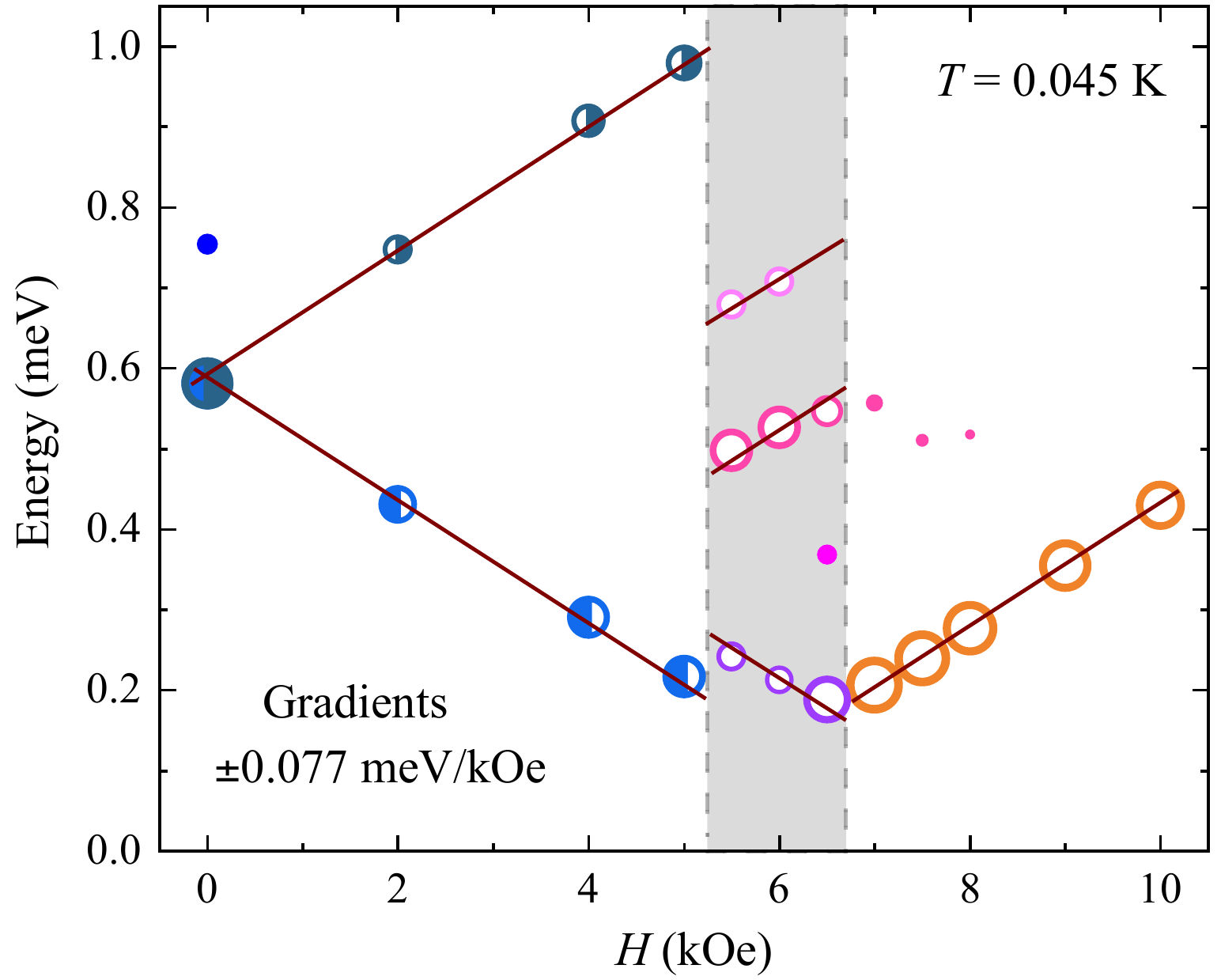}
\caption{Field dependence of the excitation energies of \esi\ at $T=0.045$~K.
(a) Cuts along the energy transfer axis of the INS spectra shown in Fig.~\ref{Fig9_IN5_maps} for $Q=[1.0,1.5]$~\AA$^{-1}$ for different applied fields.  
(b) The variation of the excitation energies with applied field.
The area of the symbols is proportional to the scattering intensity. The vertical dashed lines are guides to the eye and indicate the boundaries between the phases which we estimate to be at $H_{1}\approx 5.2$~kOe and $H_{2} \approx 6.7$~kOe.
The colouring indicates branches of excitations which follow the variation with the Zeeman energy expected in the three phases: unmagnetised ($H<H_1)$, $\nicefrac{1}{3}$-magnetised ($H_1 < H < H_2$) and fully magnetised $ H_2 \lesssim H$. 
The number of main excitations are in line with expectations in both the unmagnetised and fully magnetised phases. The lines through the points are drawn with gradients given by the effective $g$-factor reported in~\cite{Leask_1986} and assuming the excitation energy varies as $g\mu_{\rm B} H$, where $H$ is the applied field.
We do not expect the magnetisation and demagnetising field to vary significantly in each phase (see main text).}
\label{Fig10_IN5_E(H)}
\end{figure}

The field dependence of the excitations have been examined via the cuts made along the energy transfer axis in the scattering vector range 1.0 to 1.5~\AA$^{-1}$ (see Fig.~\ref{Fig10_IN5_E(H)}(a)). For clarity, the intensity cuts have been plotted separately to follow the excitations both within and outside of the plateau region.
The evolution of the excitations with field is plotted in Fig.~\ref{Fig10_IN5_E(H)}(b), with the area of each point corresponding to the intensity of the excitation branch.
The plateau phase is indicated by the dashed lines.
While demagnetisation effects prevent a direct comparison to our magnetisation measurements, the plateau region can be defined via a combination of the elastic diffraction maps accompanying the inelastic powder average spectra and examining the development of the excitations.
Although remnants of the non-integer peaks are still present in the elastic diffraction maps for 7-9~kOe, their intensities are greatly reduced.
The single, high energy excitation present in higher fields shown in orange in Fig.~\ref{Fig10_IN5_E(H)} is also present from 7~kOe, implying that the system is no longer in the plateau phase and has entered the fully magnetised state.

Distinct field-induced trends can be seen in the excitations below and above the plateau regime.
At low fields, the 0.58~meV excitation present in zero field splits into two pronounced branches.
The higher energy branch continues to increase in energy with field while that of the lower energy excitation decreases with field.
The field gradients for increasing and decreasing branches are very similar.
This can be rationalised via the spin-flipping mechanism in Ising systems for which the energy of reversing the Ising moments is varying linearly with the applied field.
At fields above the plateau, a single high-intensity excitation branch is present.
With further field increase, the energy of the excitation increases with the same gradient as the lower-field branch, implying that the system is in the fully magnetised state along the applied field.
Remnants of the 0.5~meV excitation, shown in pink in Fig.~\ref{Fig10_IN5_E(H)}(b), are present immediately above the plateau with a reduced intensity, but disappear at higher fields.

The system enters the plateau phase above 5~kOe, in this regime three intense excitation branches are observed, all varying with applied field with the same gradient.

The individual levels of the Er$^{3+}$ ions are Kramers doublets with the energy levels of the ion split by a magnetic field. This splitting is characterised by an anisotropic effective $g$-tensor. In general, this will be different for different ionic states. Extensive analysis in~\cite{Leask_1986} reported the values of the  $g$-tensor along the principal axes. The Zeeman energy of each of the Ising-like spins was well-approximated by taking $g_z=13.6$ with $g_x=g_y=0$ where $z$ is the easy-axis direction of each magnetic ion and $x$ and $y$ are the two perpendicular directions. For a field varying along the $a$~direction this would give a Zeeman energy varying as $\pm 0.077 H$~meV, illustrated by the slope of the straight lines in Fig.~\ref{Fig10_IN5_E(H)}.

\section{Modelling magnetic properties}	\label{sec_model}
\esi\ is thought to be well described as an Ising-like system.  
There are two easy magnetisation axes set by the crystal fields.
Labelling the ions in the unit cell as in Fig.~\ref{Fig3_Mag_structure}(b), ions 1 and 3 share one easy axis and ions 2 and 4 share the other.
Our Rietveld refinement gives the total moment as 6.55~$\mu_{\rm B}$ with components given in Table~\ref{tab:Positions+Moments}. 

\begin{figure}[tb] 
\includegraphics[width=0.9\columnwidth]{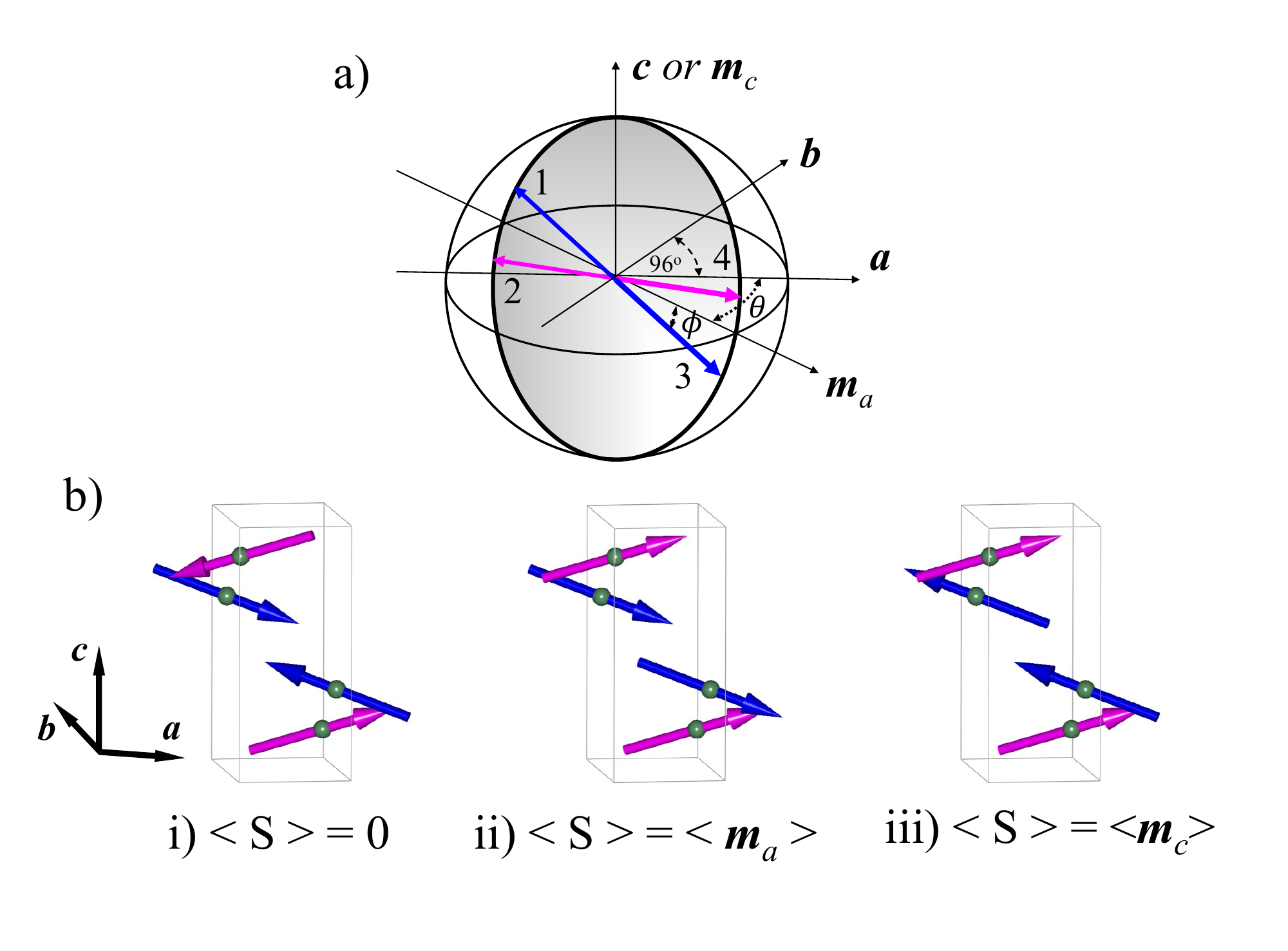}
\caption [width=0.5\columnwidth]{Configurations of the moments in possible ground states. a) Perspective view showing the spins 1 and 3 aligned along one easy axis and spins 2 and 4 aligned along the other. This corresponds to the expected ground state in zero applied field. In an applied field we expect spins to flip to maximise their alignment with the field. For a field applied parallel (or near parallel) to $\vect{a}$, we expect spins 1 and 2 to flip. This gives rise to a net magnetisation in the $\langle \vect{m}_a \rangle$ direction, which is in the plane perpendicular to the $c$ direction at an angle $\theta = 19 \degree$ to the $a$ direction. b) The orientation of the moments in the unit cell:  i) the zero-field case; ii) the fully magnetised state for a field applied in the $a$ direction. The expected average moment
per ion is $\langle \vect{m}_a\rangle$, which is given in Table \ref{tab:Positions+Moments}; and iii) the fully magnetised state for a field applied in the $c$ direction.}
\label{Fig:SpinAlignment}
\end{figure}

The angles of the average magnetisation direction $\langle \vect{m}_a \rangle$ with respect to the $a$~direction and from the $ab$-plane to the $c$ direction (see Fig.~\ref{Fig:SpinAlignment}) are $\theta=19.0 \degree$ and $\phi = 18.4 \degree$ respectively. These values differ slightly from previously reported values $\theta = 28\degree, \phi = 15 \degree$ in~\cite{Leask_1986} and $\theta = 21.3\degree, \phi = 12.8 \degree$ in~\cite{Hester_2021_D}. 
The value we obtain for $\phi$ fits with a natural explanation of the transitions observed in the magnetisation in Fig.~\ref{Fig5_Mag_a} and in Fig.~\ref{Fig6_Mag_c} as the field increases.
The jump to a magnetisation of $\sim 2 \mu_{\rm B}$ per magnetic ion for fields aligned along the $c$~direction matches the value of $2.08~\mu_{\rm B}$ of the easy-axis moment resolved along this direction. With the field aligned along the $a$~direction, the magnetisation rises quickly to 
$\gtrsim 6$~$\mu_{\rm B}$ as the field is increased from \SI{6.5}{} to $\SI{7.5}{\kilo\oersted}$. This is close to the moment $6.20$ $\mu_{\rm B}$ from our Rietveld refinement we would expect in the fully magnetised phase. The magnetisation in the centre of the intermediate phase  at \SI{0.5}{K}, which we estimate as $2.1$~$\mu_{\rm B}$ per Er ion, is close to one-third of the theoretical fully magnetised value.

 The interpretation of these transitions,  suggested by the model of \cite{Leask_1986}, is that the moments of the ions 1 and 2 reverse direction to align with the magnetic field. The ground state configuration is assumed to be changing from the one identified  as i) in Fig.~\ref{Fig:SpinAlignment} to either ii) or iii) for the case of a field along the $a$~direction or the $c$~direction. (We have labelled the ions as in Fig.~\ref{Fig3_Mag_structure}.) 
 The gradual increase of the moment for increasing fields above the transitions suggests that the easy axis slowly realigns towards the magnetic field with increasing field in both cases. The intermediate phase, with magnetisation of one third of its expected value in the fully magnetised phase ($6.55~\mu_B$), requires an enlarged unit cell in which there are two-thirds of a spin flip per original unit cell on average.

\begin{table}[tb]
\centering
\begingroup
\setlength{\tabcolsep}{6pt} 
\begin{tabular}{c  cccc c rrrr} 
 & \multicolumn{4}{c}{Ion $j$, State i)} & &\multicolumn{4}{c}{Ion $j$, State ii)} 
 \\[0.5ex]
\cline{2-5} \cline{7-10}
Ion $i$ &  1 & 2 & 3 & 4 &&  1 & 2 & 3 & 4 \\ [0.5ex]
\hline\hline
 1 & $\alpha$  & $\beta$ & $\delta$ & $\gamma$ & & $\alpha$  & $\beta$ & $-\delta$ & $-\gamma$ \\ 
2 &  $\beta$ &  $\alpha$ & $\gamma$ & $\delta$ & &   $\beta$ &  $\alpha$ & $-\gamma$ & $-\delta$ \\ 
3 & $\delta$  &  $\gamma$ &  $\alpha$ & $\beta$ & &  $-\delta$  &  $-\gamma$ &  $\alpha$ & $\beta$\\ 
4 & $\gamma$  & $\delta$ &  $\beta$ &  $\alpha$ & &  $-\gamma$  & $-\delta$ &  $\beta$ &  $\alpha$ \\ 
 \hline
 \end{tabular}
 \endgroup
 \caption{The parameters giving the contribution to the energy from interactions between an Ising-like spin of an Er ion labelled $i$ in Fig.~\ref{Fig3_Mag_structure} with  all those labelled $j$
 in the crystal. (We have used the labelling of ions from~\cite{Hester_2021_D}. The parameters $\beta$, $\gamma$ and $\delta$  were estimated in~\cite{Leask_1986} from the measured phase diagram, while the parameter $\alpha$ was estimated from spectroscopic measurements. 
 The values quoted were $\alpha = \SI{-0.088 \pm 0.006}{\milli\electronvolt}$, $\beta=\SI{0.015\pm 0.004}{\milli\electronvolt}$, $\gamma=\SI{-0.144\pm 0.004}{\milli\electronvolt}$ and $\delta=\SI{-0.050\pm 0.004}{\milli\electronvolt}$. (In \cite{Leask_1986} ion 3 is labelled 4 and vice versa.)
 }
 \label{Tab:LeaskModel}
\end{table}

The excitations seen in our inelastic data support the validity of an Ising model with two easy-axes. In any Ising system, all single spin flips are eigenstates of the Hamiltonian. Excitations are therefore local and there should be no dispersion of the excitations. As mentioned above we found no dispersion for excitations in our angle-resolved data~\cite{supplementary} and presented angle-averaged results in Fig.~\ref{Fig9_IN5_maps}, showing the existence of dispersionless bands.
In the model of~\cite{Leask_1986}, the energies of the possible ground states  are characterised by the energy of interaction of the moment $i$, with all moments of type $j$ in the rest of the lattice. The model assumes that the interactions between ions have the same symmetry as the crystal lattice.  For example, ions 1 and 3 share the same easy axis and have the same separations as ions 2 and 4. The interactions between these pairs of ions are taken to be the same, namely $\beta$. On this basis, the contribution to the ground state energy of a given ion in the unit cell with the other ions in the system in the unit cell is given in Table~\ref{Tab:LeaskModel}.

Assuming that the interactions between Ising spin variables of ions $i$ and $j$ are of the form $J(\vect{r}_i - \vect{r}_j)  S_i S_j$, where $\vect{r}_i$ is the position vector an ion $i$, the parameters in Table~\ref{Tab:LeaskModel} determine the strength of the local field acting on an individual ion. The energy of an excitation is $\epsilon_{\pm}(H) = -2(\alpha + \beta + \gamma + \delta) \pm  g\mu_{\rm B} B\cos \psi $ in state i), see Fig.~\ref{Fig:SpinAlignment}. Here $\psi$ is the angle between the magnetic field and the easy-axis for the spin. We are defining the effective $g$-factor as in \cite{Leask_1986}, so that the Zeeman energy to flip the Ising effective spin-1/2 is $g\mu_{\rm B} B\cos \psi $. 

In state i) there are two excitation energies with a splitting given by the Zeeman term. The magnetisation is small and the field $B$ is well approximated by the applied field $H$. In our experiments the field was applied along the $a$~direction and $\cos\psi = \cos \phi \cos \theta$. With the parameters quoted above and the value given for $g$ (the $g$-tensor was measured spectroscopically in \cite{Leask_1986}), the model predicts the excitation energies to be $\epsilon_\pm(H) = 0.53 \pm 0.077 H$ 
\SI{}{\milli\electronvolt}, with $H$ in \SI{}{\kilo\oersted}. This agrees remarkably well with the excitation energies shown in Figure~\ref{Fig10_IN5_E(H)} for applied fields $H < H_1 \approx \SI{5.2}{\kilo\oersted}$.
These have $\epsilon_+ \approx 0.58 + 0.077 H$ and $\epsilon_- \approx 0.58 - 0.077 H$~meV. 

We identify the ground state at applied fields $H >H_2 \approx \SI{6.7}{\kilo\oersted} $ with spin configuration ii) characterised in Fig.~\ref{Fig:SpinAlignment}.
Here the model also predicts the excitation energy:
$\epsilon(H) \approx - 2(\alpha + \beta - \gamma - \delta) + g\mu_{\rm B} H \cos \psi + C$ where $C$ is a constant. The Zeeman energy should vary linearly with the applied field as the magnetisation, and hence the demagnetising field, are expected to be constant within each phase and their effects are denoted by $C$.
There is only one excitation as any spin must flip against the field.
This implies that the excitation energy in meV is $\epsilon(H) \approx -0.24 +  g\mu_{\rm B} H \cos \psi + C $, with $H$ in \SI{}{\kilo\oersted}. The model predicts one dominant excitation.
This is close to what we find for the applied fields  $H > H_2$ shown in Fig.~\ref{Fig10_IN5_E(H)}.

In both states i) and ii), see Fig.~\ref{Fig:SpinAlignment}, there is evidence for weak scattering by other modes. In state i) there is some scattering at energy $\epsilon \approx \SI{0.75}{\milli\electronvolt}$, which we observed only in zero field. We do not have an explanation for this within the simple model we are using. In state ii) there is weak scattering at $\epsilon \sim \SI{0.5}{\milli\electronvolt}$, which disappears at higher fields. Again, we cannot explain this within the model.

The magnetic field in the sample will be significantly different from the applied field once the system magnetises.  This is because the magnetisation is not parallel to the applied field and because of demagnetising effects. For applied fields $H\lesssim 10$~kOe, the magnetisation is close to one of three values: 0, $\langle \vect{m}_a \rangle/3$ and $\langle \vect{m}_a \rangle$. The demagnetising effects are hard to quantify, but are apparent in our results. For example, the magnetisation results shown in Fig.~\ref{Fig5_Mag_a} (inset), show the transition into the fully magnetised phase occurs at an applied field $H \sim \SI{6.5}{\kilo\oersted}$ at \SI{0.5}{K}. The same transition appears at the slightly higher applied field $H_2 \sim \SI{6.7}{\kilo\oersted}$ in the sample studied in inelastic neutron scattering, see Fig~\ref{Fig10_IN5_E(H)} (where data were taken at the temperature of \SI{0.05}{K}). The samples in these two measurements were different so that direct mapping between applied fields for the magnetisation and neutron scattering measurements is not possible on account of the demagnetising effects.

In the case of the sample studied in neutron scattering, its shape was approximately a half-cylinder. The axis of the cylinder was $27 \degree$ away from the $a$~axis in the $ac$~plane. None of the applied field, the magnetisation direction $\langle \vect{m}_a\rangle$ and the demagnetising field would be aligned with the cylinder axes. For some directions of applied field the resultant field has different projections along the two easy axes, which would lead to a further splitting of the Zeeman split excitations. For the particular case of the field applied parallel to the $a$~axis, our estimates suggest that this splitting is negligible. (We assumed a uniform demagnetising tensor with component $0.15 \times 4\pi$ along the axis and equal components in the perpendicular directions.)

The identification of the intermediate phase is difficult. We would like to establish for which directions of applied field the phase exists (we have only looked along one direction) as well as find the new ordered state.  The diffraction data shown in Figs.~\ref{Fig7_WISH_Xtal} and \ref{Fig8_WISH_Xtal} indicate possible Bragg as well as diffuse peaks corresponding to propagation vectors ${\bf q} =[0\frac{1}{3}0] \mbox{ and } [0\frac{1}{4}\frac{1}{4}]$.
The nature of the phase also seems to vary with applied field. Within the easy-axis model, the magnetisation and diffraction data imply a state with enlarged periodicity in both $b$ and $c$~directions and with an average of two-thirds of a spin-flip per each original unit cell. It might be possible to explain the magnetisation with an enlarged cell, for example $4 \times 4$ but not involving a factor 3, and flipping close to two-thirds of a spin per original unit cell. However, this would make it difficult to explain the scattering corresponding to propagation vectors  ${\bf q} = [0\frac{1}{3}0]$. 

Our model assumes that the magnetisation and, hence, the demagnetising field are constant in each phase including the intermediate phase. We therefore expect only the applied component of the total field to be varying. The variation of the excitation energies with applied field would then still show the linear dependence of the Zeeman energy seen in the unmagnetised and fully magnetised cases. This is close to what we see. The small deviation from this behavior, slightly below the transition at $H = H_2$ (see Figure~\ref{Fig10_IN5_E(H)}), may also be connected with the variation of the phase seen in the diffraction data.

\section{Summary}
Magnetic measurements and a range of neutron scattering techniques have been employed to investigate the in-field behaviour of D-type \esi\, an Ising-like antiferromagnet.
Powder neutron diffraction data confirm the formation of a four-sublattice ${\bf q}=0$ structure below the ordering temperature of $T_{\rm N}=1.85$~K with two pairs of collinear \afm\ moments within the crystallographic unit cell. 
Magnetisation measurements with the field applied along the $a$~axis of a single crystal of \esi\ reveal the existence of a narrow and temperature-sensitive magnetisation plateau at $\frac{1}{3}$ of the saturation magnetisation.
In-field single crystal neutron diffraction has shown that the magnetic unit cell undergoes a significant increase within the plateau regime demonstrated by the appearance of scattering signal at non-integer positions described by the propagation vectors ${\bf q} = (0 \frac{1}{2} \frac{1}{2})$, ${\bf q} = (0 \frac{1}{4} \frac{1}{4})$ and ${\bf q} = (0 \frac{1}{3} 0)$.
The intensity, shape and width of these field-induced features change rapidly with the applied magnetic field within the magnetisation plateau.   
The diffraction pattern then returns to integer peaks at higher fields, upon its entrance into the paramagnetic phase and out of the plateau.

Inelastic neutron scattering data indicate flat, dispersionless excitations, as expected in an Ising-like system.
Below the plateau, two intense excitation branches are observed, one linearly increasing with field and the other decreasing while above the plateau, only one branch is visible.
The field gradients for all three branches are well described by the previously determined $g$-factor for the Er$^{3+}$ magnetic ions.
Within the plateau regime, the excitation spectrum corresponds to an increased magnetic unit cell and contains three intense branches. This observation should restrict the number of candidate ground states significantly. Enlarging the magnetic unit cell leads quickly to an increase in the number of inequivalent spin flips and hence the number of excitations, while our data show only three significant excitations.

\section{ACKNOWLEDGMENTS}
We acknowledge the technical support and expertise of the sample environment teams at ISIS and ILL. The work at the University of Warwick was supported by EPSRC through grants EP/M028771/1 and EP/T005963/1.
\bibliography{RE2Si2O7_all}
\end{document}


\title{{\it Supplementary Material for} \\ Magnetic structure, excitations and field induced transitions in the honeycomb lattice $\mathbf{Er_2Si_2O_7}$ }
\author{M.~Islam}			\email{Manisha.Islam@warwick.ac.uk} \affiliation{Department of Physics, University of Warwick, Coventry CV4 7AL, United Kingdom}
\author{N.~d'Ambrumenil}		\affiliation{Department of Physics, University of Warwick, Coventry CV4 7AL, United Kingdom}
\author{D.~D.~Khalyavin} 
\affiliation{ISIS Facility, STFC Rutherford Appleton Laboratory, Chilton, Didcot, OX11 0QX, United Kingdom}
\author{P.~Manuel}
\affiliation{ISIS Facility, STFC Rutherford Appleton Laboratory, Chilton, Didcot, OX11 0QX, United Kingdom}
\author{F.~Orlandi}
\affiliation{ISIS Facility, STFC Rutherford Appleton Laboratory, Chilton, Didcot, OX11 0QX, United Kingdom}
\author{J.~Ollivier}			\affiliation{Institut Laue-Langevin, 71 Avenue des Martyrs, CS 20156, 38042 Grenoble Cedex 9, France}
\author{M.~Ciomaga Hatnean}	\affiliation{Department of Physics, University of Warwick, Coventry CV4 7AL, United Kingdom}
\altaffiliation[Current affiliations: ]{Materials Discovery Laboratory, Department of Materials, ETH Zurich, 8093 Zurich  and Laboratory for Multiscale materials eXperiments, Paul Scherrer Institute (PSI), 5232 Villigen, Switzerland}
\author{G.~Balakrishnan}		\affiliation{Department of Physics, University of Warwick, Coventry CV4 7AL, United Kingdom}
\author{O.~A.~Petrenko}		\email{O.Petrenko@warwick.ac.uk} \affiliation{Department of Physics, University of Warwick, Coventry CV4 7AL, United Kingdom}

\date{\today}
\maketitle
\section*{Single Crystal Neutron Diffraction on WISH}

\begin{figure*}[b] 
\includegraphics[width=\textwidth]{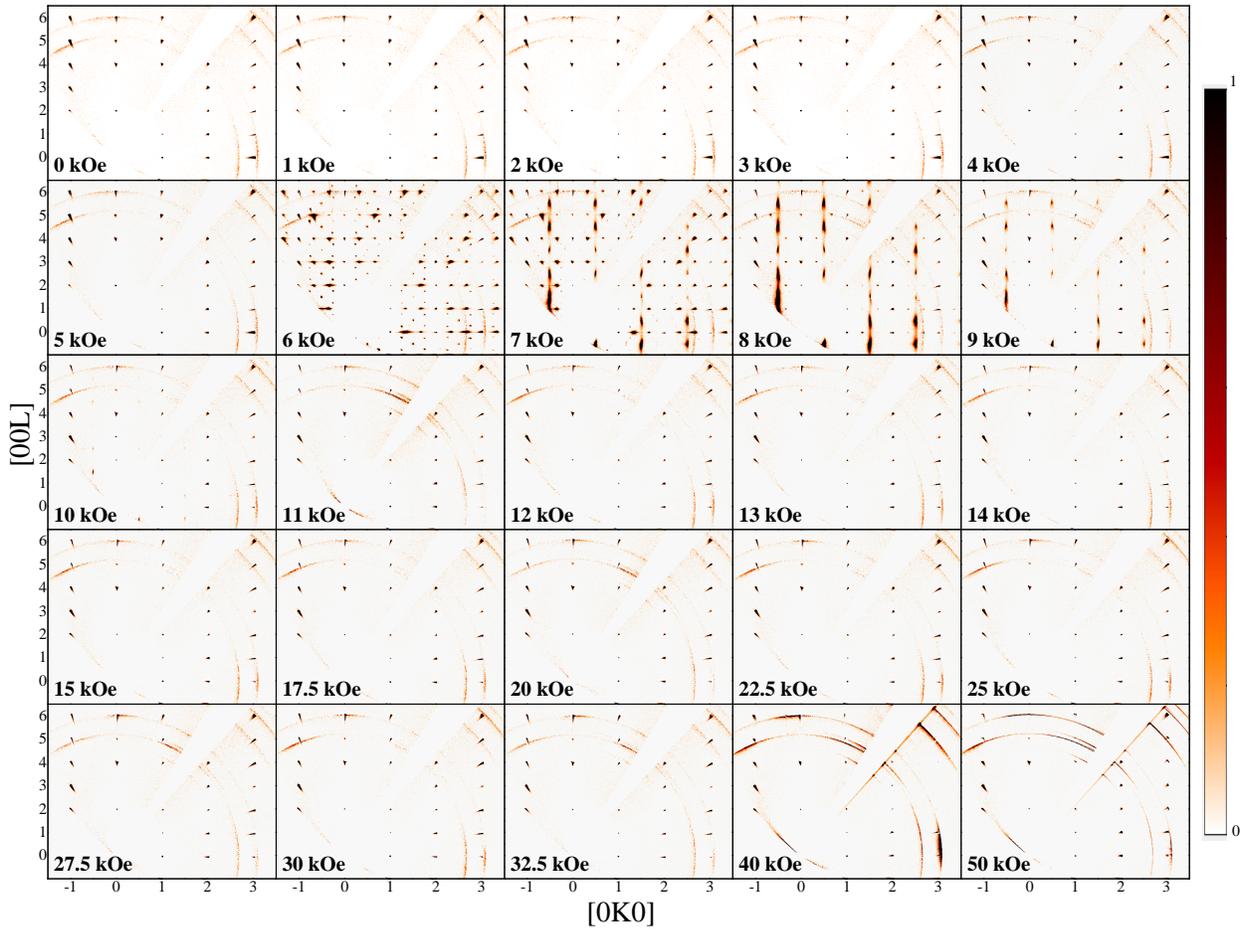}
\caption [width=2\columnwidth]{Magnetic single crystal neutron diffraction maps of the $(0kl)$ plane for D-type \esi\ measured on the WISH diffractometer at $T=1.5$~K in an increasing field applied along the $a$~axis. The magnetic signal is obtained by subtracting the $T=2$~K background. Sample movement at $H=40$~kOe and $H=50$~kOe resulted in misalignment and was corrected for using separate UB matrices.} 
\label{Fig12_WISH_supplementary_up}
\end{figure*}

The detailed measurements of the magnetisation process for a single crystal of D-type \esi\ on the WISH diffractometer at ISIS are shown in SF.~\ref{Fig12_WISH_supplementary_up}, for an increasing field applied along the $a$ axis. The magnetic signal was isolated via the subtraction of the higher temperature nuclear signal at T~=~2~K.
The observed magnetic diffraction pattern corresponds to a q~=~0 magnetic structure in applied fields above and below the magnetic plateau region. Additional diffuse signal appears on and around the plateau at non-integer positions in reciprocal space. The reciprocal space consists of only Bragg peaks upon leaving the plateau phase at 10~kOe. 

\begin{figure*}[tb] 
\includegraphics[width=\textwidth]{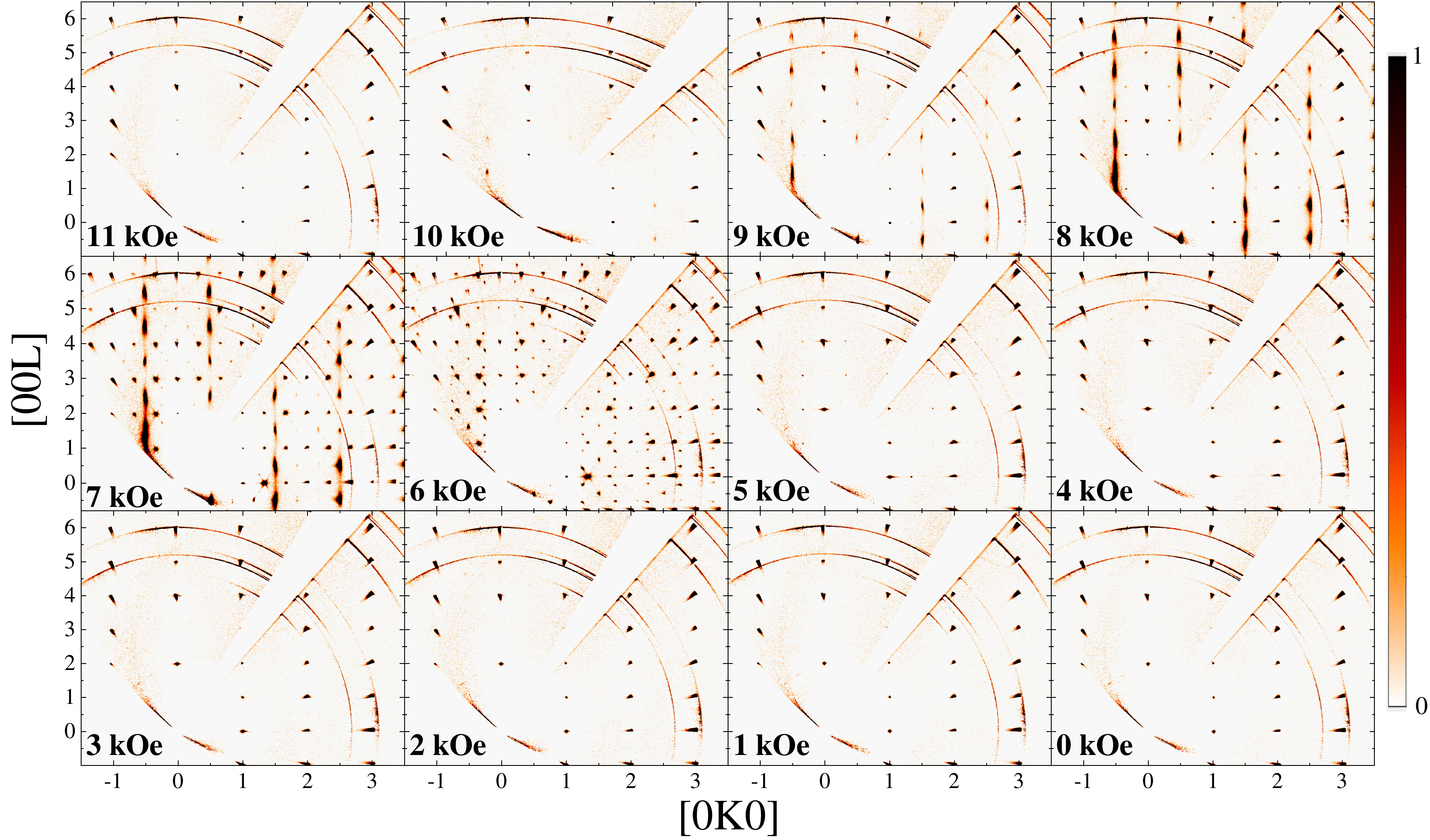}
\caption [width=2\columnwidth]{Magnetic single crystal neutron diffraction maps of the $(0kl)$ plane for D-type \esi\ measured on the WISH diffractometer at $T=1.5$~K in a decreasing field applied along the $a$~axis. The magnetic signal is obtained by subtracting the $T=2$~K background. The appearance of distinct powder rings are a result of sample movement and was corrected for using separate UB matrices.}
\label{Fig13_WISH_supplementary_down}
\end{figure*}

Powder rings are visible at 40~kOe and 50~kOe due to sample movement and resulting misalignment of the sample. This has been corrected for using a new UB matrix. This correction has also been made in SF.~\ref{Fig13_WISH_supplementary_down}, in which the reciprocal maps for a decreasing field are presented. The same diffuse features are again seen for fields within the plateau regime. 
\section*{Single Crystal Inelastic Neutron Scattering on IN5}
\begin{figure*}[tb] 
\includegraphics[width=\textwidth]{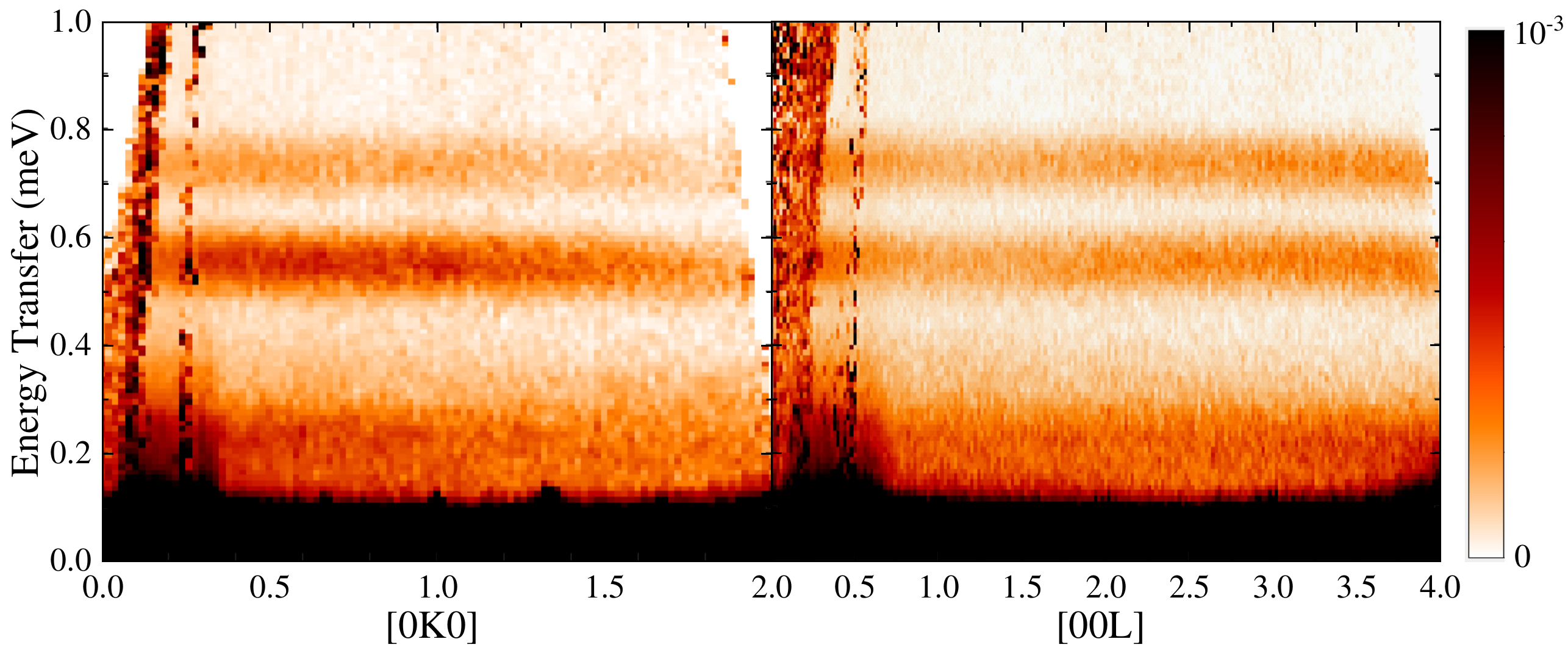}
\caption [width=2\columnwidth]{Example of inelastic neutron scattering measurements along [0K0] and [00L] of a single crystal sample of D-type \esi\ at $T=0.045$~K in 6~kOe, taken on the IN5 spectrometer. The sample was rotated along the direction of the applied field, parallel to $a$. The high intensity features at  K$=0.25$ and L$=0.25$ are artefacts of the cryomagnet and can be ignored. The excitations are flat and dispersionless in both directions.}
\label{IN5_supplementary}
\end{figure*}

\begin{figure*}[tb] 
\includegraphics[width=\textwidth]{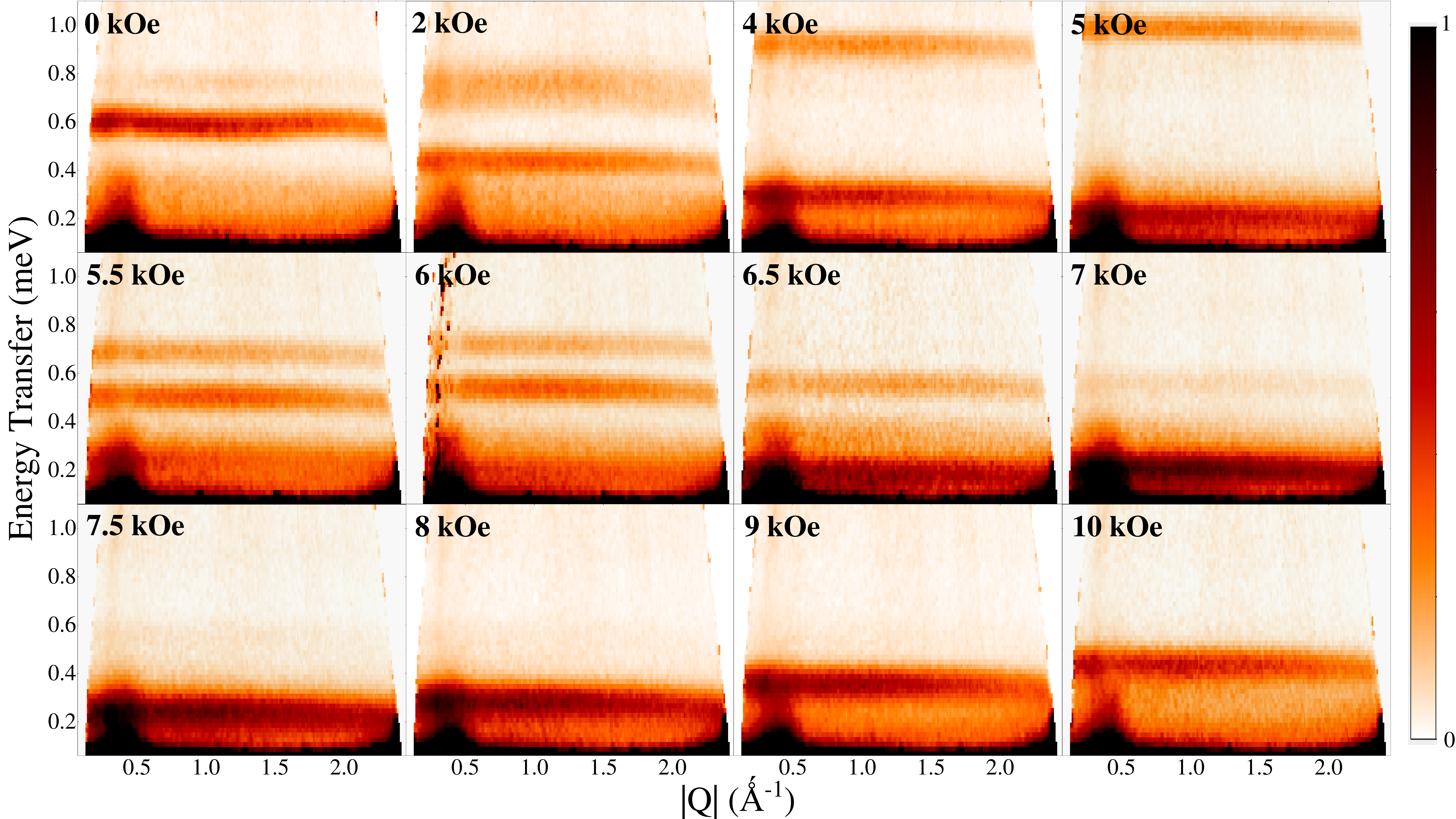}
\caption [width=2\columnwidth]{Full set of powder-averaged energy spectra measured using the IN5 spectrometer on a single crystal sample of D-type \esi\ at $T=0.045$~K in different fields. The sample was rotated along the direction of an applied field, $H \parallel a$, while the powder average is taken in the $b^\star$-$c^\star$ scattering plane. The high intensity feature at $|Q|=0.35~\rm{\AA^{-1}}$ is an artefact of the cryomagnet and can be ignored.}
\label{Fig14_IN5_supplementary_powder_average}
\end{figure*}

The in-field excitation spectra of \esi\ is shown in SF.~\ref{Fig14_IN5_supplementary_powder_average}. Low energy excitation measurements were performed on the IN5 spectrometer at ILL at 0.045~K with the field applied along the $a$ axis. As shown in the data taken in 6~kOe in SF.~\ref{IN5_supplementary}, only dispersionless excitations were observed with no significant dependence of the intensity on the scattering vector. We have therefore presented our inelastic measurements as an average over the rotation of the crystal$\sim43~^{\circ}$ and grouping the pixellated data from the IN5 detector as (Q, E) bins, or Debeye-Scherrer cones, over the entire detector range.  This produces a powder-average. However, this method cannot produce a true powder-average as all orientations of the crystal, with incident and scattered neutrons spanning all possible crystal symmetries, have not been measured. The full set of data presented in SF.~\ref{Fig14_IN5_supplementary_powder_average} represents the average for the directions of interest, as the crystal has been oriented for measurement in the $b^\star$-$c^\star$ scattering plane. 

\begin{figure*}[tb] 
\includegraphics[width=\textwidth]{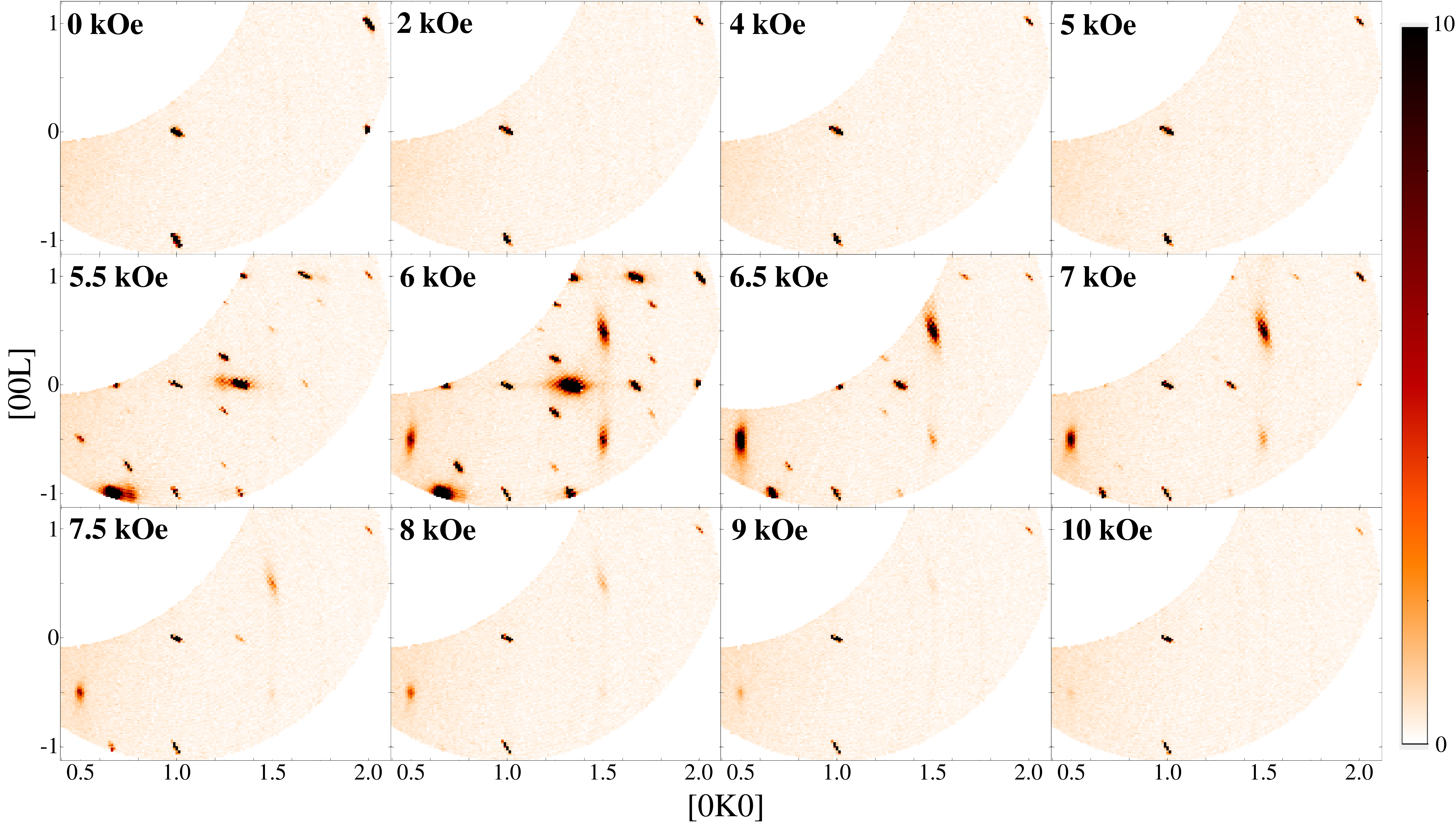}
\caption [width=2\columnwidth]{Full set of magnetic single crystal neutron diffraction maps for zero-energy transfer of the $(0kl)$ plane for D-type \esi\ measured on IN5 at $T=0.045$~K in fields applied along the $a$~axis.}
\label{Fig15_IN5_supplementary_elastic_maps}
\end{figure*}

The elastic reciprocal space maps in SF.~\ref{Fig15_IN5_supplementary_elastic_maps} are presented to accompany the inelastic spectra in order to identify the magnetic regime of the system. The system enters the plateau at 5.5~kOe with the appearance of the non-integer magnetic peaks. The plateau ends at 7.5~kOe, however, some low intensity non-integer peaks remain until 10~kOe. 

\begin{figure*}[tb] 
\includegraphics[width=\textwidth]{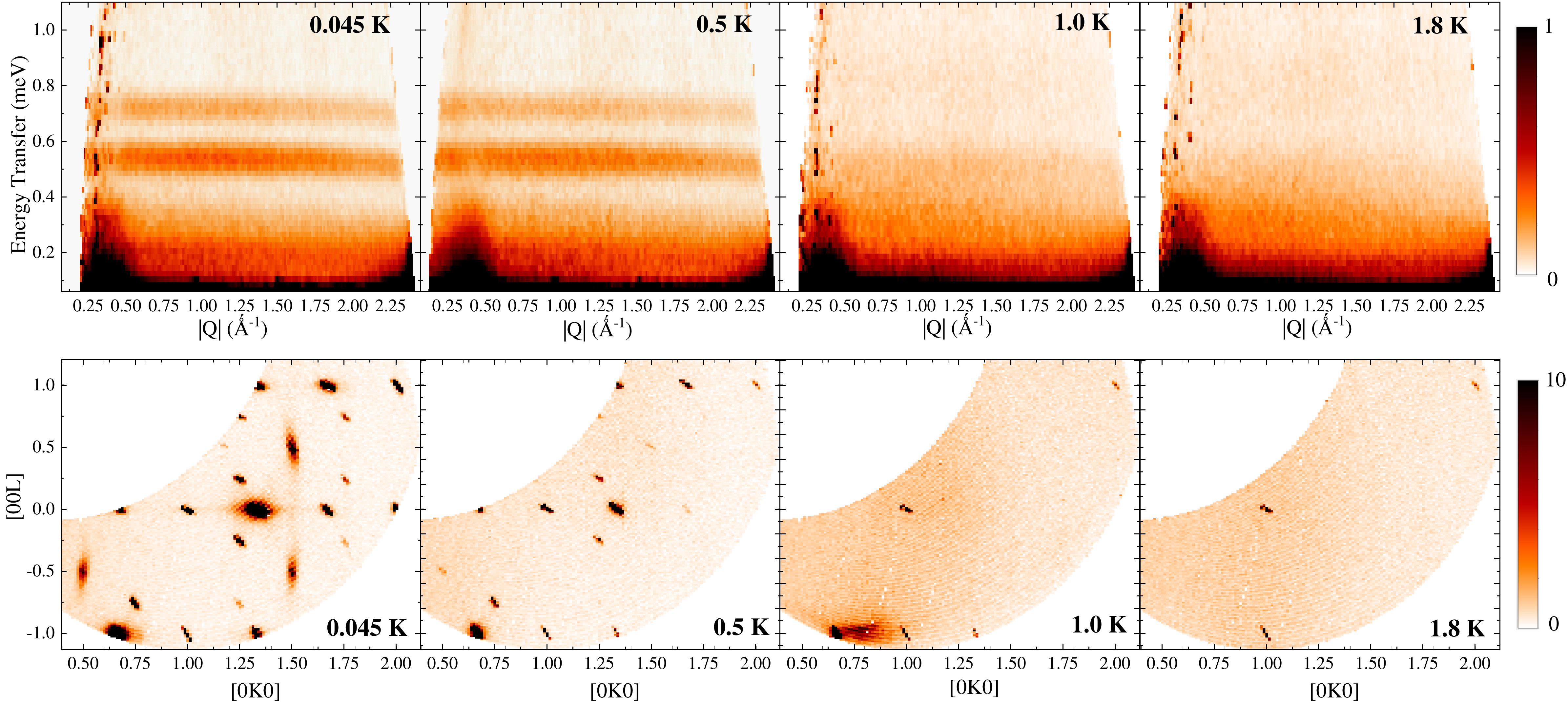}
\caption [width=2\columnwidth]{(top) Powder-averaged energy spectra measured using the IN5 spectrometer on a single crystal sample of D-type \esi\ at $H=6~kOe$ applied along the $a$~axis. The sample was rotated along the direction of an applied field, $H \parallel a$, while the powder average is taken in the $b^\star$-$c^\star$ scattering plane. The high intensity feature at $|Q|=0.35~\rm{\AA^{-1}}$ is an artefact of the cryomagnet and can be ignored. (bottom) Corresponding diffraction maps for zero-energy transfer in the $(0kl)$ plane.}
\label{Fig16_IN5_supplementary_temperature}
\end{figure*}

The temperature dependence of the plateau phase is presented in SF.~\ref{Fig16_IN5_supplementary_temperature} at an applied field of 6~kOe. Three distinct excitations have been observed at 0.045~K at energies of 0.2, 0.55 and 0.75~meV. The same excitations are observed at 0.5~K, despite a significant change in the corresponding elastic reciprocal space map, in which the peaks at postion ${\bf q} = (0 \frac{1}{3} \frac{1}{2})$ reduce in size and intensity. At 1.0~K, the excitations disappear and only a new non-integer peak of scattering vector ${\bf q} = (0 0.8 0)$ position is present, suggesting a new phase transition. Only integer Bragg peaks remain at 1.8~K, below the ordering temperature of $T_{\rm N}=1.85$~K. The loss of these low energy excitations with increasing temperature indicate the presence of crystal fields. 

\bibliography{RE2Si2O7_all}